\documentclass[12pt,preprint]{aastex}

\newcommand{\Lsun}{\mbox{$L_{\sun}$}}

\shorttitle{High-J CO in Galaxies}
\shortauthors{Mashian et al.}
\usepackage{graphicx}
\usepackage{amsfonts}
\usepackage{placeins}
\begin{document}

   \title{High-J CO SLEDs in nearby infrared bright galaxies\\ observed by {\it Herschel}\thanks{{\it Herschel} is an ESA space observatory with science instruments provided by European-led
Principal Investigator consortia and with important participation from NASA.} - PACS}

\author{
N. Mashian\altaffilmark{1,3},
E. Sturm\altaffilmark{2},
A. Sternberg\altaffilmark{3},
A. Janssen\altaffilmark{2},
S. Hailey-Dunsheath\altaffilmark{4},
J. Fischer\altaffilmark{5},
A. Contursi\altaffilmark{2},
E. Gonz{\'a}lez-Alfonso\altaffilmark{6},
J. Graci{\'a}-Carpio\altaffilmark{2},
A. Poglitsch\altaffilmark{2},
S. Veilleux\altaffilmark{7}
R. Davies\altaffilmark{2},
R. Genzel\altaffilmark{2},
D. Lutz\altaffilmark{2},
L. Tacconi\altaffilmark{2},
A. Verma\altaffilmark{8},
A. Wei{\ss}\altaffilmark{9},
E. Polisensky\altaffilmark{5},
T. Nikola\altaffilmark{10}
}

\altaffiltext{1}{Harvard-Smithsonian Center for Astrophysics, 60 Garden
Street, Cambridge, MA 02138, USA}
\altaffiltext{2}{Max-Planck-Institute for Extraterrestrial Physics
(MPE), Giessenbachstra{\ss}e 1, 85748 Garching, Germany}
\altaffiltext{3}{The Raymond and Beverly Sackler School of Physics and
Astronomy, Tel Aviv University, Tel Aviv 69978, Israel}
\altaffiltext{4}{California Institute of Technology, 1200 E California
Blvd, Pasadena CA 91125, USA}
\altaffiltext{5}{Naval Research Laboratory, Remote Sensing Division,
4555 Overlook Ave SW, Washington, DC 20375, USA}
\altaffiltext{6}{Universidad de Alcal{\'a} de Henares, 28871 Alcal{\'a}
de Henares, Madrid, Spain}
\altaffiltext{7}{Department of Astronomy, University of Maryland,
College Park, MD 20742, USA}
\altaffiltext{8}{Sub-dept. of Astrophysics, Denys Wilkinson Building, University of Oxford, Keble Road, Oxford, OX1 3RH, UK}
\altaffiltext{9}{Max-Planck-Institute for Radioastronomy (MPIfR), Auf
dem H\"ugel 69, 53121 Bonn, Germany}
\altaffiltext{10}{Cornell University, Ithaca, NY 14853, USA }
\email{nmashian@physics.harvard.edu}


\begin{abstract}
{We report the detection of far-infrared (FIR) CO rotational emission from nearby active galactic nuclei (AGN) and starburst galaxies, as well as several merging systems and Ultra-Luminous Infrared Galaxies (ULIRGs). Using Herschel-PACS, we have detected transitions in the J$_{upp}$ = 14 - 20 range ($\lambda \sim$ 130 - 185 $\mu$m, $\nu \sim$ 1612 - 2300 GHz) with upper limits on (and in two cases, detections of) CO line fluxes up to J$_{upp}$ = 30. The PACS CO data obtained here provide the first well-sampled FIR extragalactic CO Spectral Line Energy Distributions (SLEDs) for this range, and will be an essential reference for future high redshift studies.
We find a large range in the overall SLED shape, even amongst galaxies of similar type, demonstrating the uncertainties in relying solely on high-J CO diagnostics to characterize the excitation source of a galaxy.
 Combining our data with low-J line intensities taken from the literature, we present a CO ratio-ratio diagram and discuss its potential diagnostic value in distinguishing excitation sources and physical properties of the molecular gas.
The position of a galaxy on such a diagram is less a signature of its excitation mechanism, than an indicator of the presence (or absence) of warm, dense molecular gas.
  We then quantitatively analyze the CO emission from a subset of the detected sources with Large Velocity Gradient (LVG) radiative transfer models to fit the CO SLEDs. Using both single-component and two-component LVG models to fit the kinetic temperature, velocity gradient, number density and column density of the gas, we derive the molecular gas mass and the corresponding CO-to-H$_2$ conversion factor, $\alpha_{CO}$, for each respective source.
For the ULIRGs we find $\alpha$ values in the canonical range 0.4 - 5 M$_\odot$/(K kms$^{-1}$pc$^2$), while for the other objects, $\alpha$ varies between 0.2 and 14.}
  Finally, we compare our best-fit LVG model results with those obtained in previous studies of the same galaxies and comment on any differences. 
\end{abstract}


   \maketitle
%

\section{Introduction}

Molecular spectral line energy distributions (SLEDs) provide us with the opportunity to probe the average state of the molecular gas in galaxies and estimate the total and star-forming molecular gas masses in these sources. Carbon monoxide (CO), the most abundant molecule after molecular hydrogen, is the most commonly employed  tracer of interstellar molecular gas. With the small gaps between its energy levels allowing for a fine sampling of density-temperature relations, CO has become an important tool in the study of the star formation energetics and the effects of AGN in nearby galaxies.

The lowest 3 rotational transitions of CO, which trace the cooler gas component, are relatively easily accessible with ground-based radio and submillimeter telescopes, and have been observed in many local galaxies.
On the other hand, far-IR CO rotational lines, with J$_{upp}$ $\geq$ 13, arise from states 500 - 7000 K above ground and have critical densities of 10$^6$ $\sim$ 10$^8$ cm$^{-3}$ \citep{2012ApJ...755...57H}. These lines, which trace the warmer, denser molecular gas in the center of galaxies, are difficult to excite solely with star formation and can thus be used to test models that distinguish between AGN and starburst systems. However, until the advent of Herschel, the diagnostic use of these higher rotational levels was poorly developed since these lines were either difficult to observe or completely inaccessible from the ground.

The Herschel Space Observatory \citep{2010A&A...518L...1P} is uniquely suited to measure the submillimeter properties of nearby galaxies in a frequency range that cannot be observed from the ground.
This paper focuses on the data obtained with the Photodetector Array Camera and Spectrometer (PACS) \citep{2010A&A...518L...2P} on-board the Herschel Space Observatory, which provides observations in the 60 - 210 $\mu$m wavelength range, sampling the FIR CO emission of high-J transition lines between J$_{upp}$ = 14 up to J$_{upp}$ = 50.

Complementary CO observations are provided by the Spectral and Photometric Imaging Receiver (SPIRE) \citep{2010A&A...518L...3G} which consists of an imaging Fourier Transform Spectrometer (FTS) continuously covering the spectral range from 190 - 670 $\mu$m. The CO rotational ladders (from J$_{upp}$ = 4 to J$_{upp}$ = 13) of several extragalactic sources have been studied with SPIRE-FTS within the last five years, including the nearby starbursts and AGNs NGC\,253, M\,82, IC\,694, and NGC\,3690 \citep{2014A&A...564A.126R,2010A&A...518L..37P,2014arXiv1407.2055R}, and the local ULIRGs Arp\,220, NGC\,6240, and Mrk\,231 \citep{2011ApJ...743...94R, 2013ApJ...762L..16M, 2010AA...518L..42V}.
In \citet{2012ApJ...755...57H} we have carried out a detailed analysis of the full CO SLED of the Seyfert 2 galaxy NGC\,1068
using ground based and PACS observations, which was complemented by an analysis of the SPIRE data for this object
by \citet{2012ApJ...758..108S}.

In this paper,
we want to address the question of to what extent high-J CO transitions can be used to trace the excitation conditions in galactic nuclei, in particular the existence and physical properties of starburst-heated, AGN-heated or shock-heated gas.
We therefore present the PACS $^{12}$CO line dataset for a sample of local sources including nearby starburst galaxies (NGC\,253, M\,83, M\,82, IC\,694, NGC\,3690), Seyfert galaxies (NGC\,4945,  Circinus, NGC\,1068, Cen\,A), (U)LIRGs (NGC\,4418, Arp\,220, NGC\,6240, Mrk\,231), and the nearby interacting system (Antennae). In section 2 we describe the source sample and the Herschel-PACS observations of the FIR CO lines in these sources, as well as observations of the CO(18-17) and CO(20-19) line fluxes and upper limits of 19 ULIRGs of the Revised Bright Galaxy Sample (RBGS). In section 3 we discuss the potential use of line ratios as a diagnostic tool for distinguishing between different energy sources responsible for gas excitation, e.g. starbursts or AGN, and introduce the LVG radiative transfer modeling technique used to analyze the full CO SLEDs. In section 4 we present estimates of the physical parameters characterizing the molecular gas, obtained from single-component and two-component LVG fits to the data. We also derive the CO-to-H$_2$ conversion factors in these sources and compare our LVG results to those obtained in previous studies of these galaxies. We conclude with a summary of our findings and their implications in section 5.

\section{Target Selection, Observations and Data Reduction}
\label{sect:observations}

The {\it Herschel} data presented here are part of the guaranteed time key program SHINING (Survey with Herschel of the ISM in Nearby INfrared Galaxies, PI: E. Sturm), which studies the far-infrared properties of the ISM in
starbursts, Seyfert galaxies and infrared luminous galaxies, as well as the OT1 and OT2 follow up programmes (PIs: S. Hailey-Dunsheath and J. Fischer) and the OT1 programme of R. Meijerink.
The observations were made with the PACS spectrometer on board the Herschel Space Observatory.

To fully characterize the high-J CO emission associated with star formation, AGN, and large-scale shocks in galaxies, we defined a sample of objects consisting of 4 starbursts (NGC\,253, M\,83, M\,82, IC\,694),
4 Seyferts (NGC\,4945, Circinus, NGC\,1068,  Cen\,A), three ULIRGs (Arp\,220, NGC\,6240, Mrk\,231), the highly obscured LIRG NGC\,4418, and the nearby, well-studied interacting system NGC\,4038/4039 (Antennae). The observations of IC\,694 were complemented by observations of NGC\,3690 (a mixed source with a low luminosity AGN). Together, these two objects are known as the merger system Arp\,299.
For the Antennae system we obtained 2 pointings, one centered on the nucleus of NGC\,4039 and one on the overlap region between NGC\,4038 and NGC\,4039. The AGN subsample is restricted to the most nearby systems, where the high spatial resolution of PACS ($\sim$ 10 arcsec corresponding to $\sim$ 200 pc at $d$ = 4 Mpc) helps separate the AGN-heated gas from most of the gas in the circumnuclear star-forming region.

In most of these galaxies, we measured a set
of 7 far-IR CO transitions to provide a coarse
but sensitive sampling of the CO SLED over the full FIR range:
 CO(15-14), CO(16-15), CO(18-17), CO(20-19), CO(22-21), CO(24-23), and CO(30-29).
We have carefully inspected our full range (55-200 $\mu$m) scan of Arp\,220 \citep{2012A&A...541A...4G}, which is rich in molecular lines, and verified that these CO lines are not contaminated by other molecular features.
For some of the objects, additional PACS high-J CO data are available: with SHINING we obtained
full range PACS spectra of NGC\,1068, NGC\,4418, NGC\,4945, and M\,82, together
with full spectra of Mrk\,231 and NGC\,6240 from a SHINING-related OT2 programme (PI: J. Fischer). These data cover  high-J transitions between J$_{upp}$ = 14 and J$_{upp}$ = 50.
In the case of NGC\,6240,
the CO(16-15), CO(18-17), CO(24-23) data were taken from the shared OT1 programmes of R. Meijerink and S. Hailey-Dunsheath;  the CO(14-13) and CO(28-27) lines of NGC\,6240 and IC\,694 were taken from the Meijerink OT1 programme as well.
Combined with existing ground-based and SPIRE observations, these
measurements sample the CO ladder from CO(1-0) to CO(30-29) in a number of well-studied galaxies that are often used as templates, allowing us to address the questions posed in the introduction, e.g. to what extent high-J CO transitions can be used to unambiguously trace the existence and physical properties of AGN-heated gas.

As part of the SHINING observations, we also obtained CO(18-17) and CO(20-19) fluxes or upper limits
in 19 ULIRGs of the RBGS \citep{2003AJ....126.1607S}, in addition to Arp\,220, Mrk\,231 and NGC\,6240.
These are local ULIRGs ($z$ $<$ 0.1) with total 60 $\mu$m flux density greater than 5.24 Jy (and L$_{IR}>$10$^{12}\Lsun$).
We include these data in our study below in order to assess the use of high-J CO transitions
as probes of the excitation mechanisms in luminous dusty sources.
The observation details for the PACS data, together with galaxy classifications, are summarized in
Table~\ref{table:OBSID}.

Finally, for a subset of the RBGS ULIRGs for which no CO(6-5) data existed in the literature,
we observed this line using APEX\footnote{This part of the publication is based on data acquired with the
Atacama Pathfinder Experiment (APEX). APEX is a collaboration between
the Max-Planck-Institut fur Radioastronomie, the European Southern
Observatory, and the Onsala Space Observatory.} with the CHAMP$^+$ receiver (see Table~\ref{table:RBGS}). The APEX data were reduced with the standard software in CLASS. Calibration was obtained with the APEX calibration software \citep{2006A&A...454L..25M}.

Most of the PACS CO observations were made in PACS range scan mode with a 2600 km s$^{-1}$ velocity
coverage, while for the merging systems we increased this to 3000 km s$^{-1}$ to ensure the detection of broad lines. The PACS data reduction
was done using the standard PACS reduction and
calibration pipeline (ipipe) included in HIPE 6.
For the final calibration we normalized the
spectra to the telescope flux and recalibrated it with a
reference telescope spectrum obtained from dedicated
Neptune continuum observations. This `background normalization' method is described in the PACS data reduction guide. We also assessed the line flux uncertainties associated with the uncertainties in defining the continuum and estimated an absolute line flux accuracy of 20\% with this method.

The PACS spectrometer performs integral field spectroscopy over a 47'' $\times$ 47'' FoV, resolved into a 5 $\times$ 5 array of 9.4'' spatial pixels (spaxels.) With the exception of M\,83, M\,82, and NGC\,3690, the high-J lines in the PACS range for the remaining sources are all consistent with arising in the central spaxel, with little, if any, flux detected outside of the central spaxel. The fluxes  were thus extracted from the central spaxel only ($\theta$ = 9.4"), and referenced to a point source by dividing by the recommended point source correction factors as given in the PACS manual. We did spot checks with HIPE 11 and verified that this does not yield significantly different line fluxes or upper limits.

 In the case of M\,83 and M\,82 where the high-J CO emission are extended beyond the central spaxel, the fluxes were derived by integrating over the entire PACS FoV. For NGC\,3690, the PACS flux measurements cover both the galaxy, as well as an extended region of star formation where the galaxy disks of NGC\,3690 and IC\,694 overlap. The high-J line fluxes for this source were thus derived by co-adding the 3 spaxels covering components B and C (as NGC\,3690 and the overlap region are respectively referred to in the literature).

To supplement these PACS line observations, we collect low and mid-J line fluxes for these sources from the literature and, when necessary, apply aperture corrections using available CO maps.
For point-like sources such as Arp\,220, NGC\,6240, and Mrk\,231, where the low and mid-J line fluxes are fully contained in the PACS beam, no aperture corrections are necessary. This is also deemed to be the case for NGC\,4418 and IC\,694, where $^{12}$CO J=1-0, J=2-1, and J=3-2 maps reveal the emission to be fully enclosed within a $\theta\sim$ 10" beam \citep{2013ApJ...764...42S, 2012ApJ...753...46S,1999A&A...346..663C}.
The morphology of NGC\,1068, which is composed of a compact central circumnuclear disk ($\theta \sim$ 4") and an extended ring ($\theta \sim$ 20-40''), is slightly more complex. To avoid potential scaling issues, we only consider lines that are probed on the scale of the compact source at the center of NGC\,1068, i.e. the interferometric measurements of the  J$_{upp}$ = 1 - 3 lines integrated over the central 4'' by \citet{2011ApJ...736...37K}, and the J$_{upp}$ = 9 - 13 lines probed by SPIRE FTS with a beam width of $\theta \sim$ 17'' \citep{2012ApJ...758..108S}.

In the case of M\,83, aperture corrections are also deemed unnecessary since PACS flux measurements for this source extend over a $\sim$ 21'' area, and the low and mid-J lines found in the literature refer to a similar beam size \citep{2006AA...460..467B, 2001AA...371..433I}. Similarly, for M\,82, the J$_{upp}$ = 4 - 13 $^{12}$CO lines are referenced to the $\sim$ 43'' beam size of the SPIRE spectrometer \citep{2012ApJ...753...70K} and therefore do not need to be scaled when compared against the high-J lines which fill the PACS FoV ($\theta$ = 47'') in this source. The $^{12}$CO J=1-0 map for M\,82 indicates that the low-J lines are also not in need of any scaling corrections since the emission is predominately contained within an $\sim$25'' area, consistent with the FWHP beam of 24.4'' referenced for the J$_{upp}$ = 1 - 3 line measurements  \citep{2003ApJ...587..171W}.

 $^{12}$CO J=1-0, J=2-1, and J=3-2 maps for the Arp\,299 merger system show that the source sizes of NGC\,3690 and the overlap region are $\theta \sim$ 2'' and 4'' respectively, and that the combined emission from these two components does not extend beyond $\theta \sim$ 19'' (the PACS aperture size over which the high-J lines were integrated) \citep{2012ApJ...753...46S}. The low-J line fluxes for this region are thus derived by co-adding the flux values presented in the literature for components B and C without any further scaling corrections. The mid-J line fluxes are taken from SPIRE observations which include contributions from both components, B and C \citep{2014arXiv1407.2055R}. Although the SPIRE beam sizes vary (15-42''), it is assumed that the fluxes, too, are mostly contained within a 19'' beam and no further scaling is applied.

In the case of NGC\,4945 and Circinus, available CO maps of the low-J lines indicate that these sources are spatially resolved and more extended than the PACS single spaxel size \citep{1993A&A...270...29D,1996A&A...309..705M,2008MNRAS.389...63C,2014arXiv1407.1444Z}. Since the recorded fluxes fill the respective beams with which they were probed, we assume a uniform distribution of flux and linearly scale the low-J emission with the ratio of the PACS spaxel area and the respective literature beam area. A similar technique is applied to NGC\,253, where the extended low and mid-J lines corrected to a 15'' beam are assumed to be uniformly distributed over that beam size and are thus scaled linearly with the PACS area \citep{2014A&A...564A.126R, 2008ApJ...689L.109H}.

The PACS measurements, along with the (aperture-corrected) low-J line fluxes collected from the literature, are summarized in Table~\ref{table:COlines}.

\section {Excitation Analysis}

\subsection{CO SLEDs and Line Ratios}

The observed CO SLEDs of the sources, whose excitation conditions will further be explored in the following sections, display a large variation in overall SLED shapes (Figure~\ref{fig:CO_SEDs}). On a first glance
this spread seems to nicely follow the qualitatively expected behavior that the strength of high-J CO lines increases with increasing importance of an AGN in these sources. For instance, the CO SLED of the archetypal starburst galaxy M\,82 peaks at around J$_{upp}$ = 7 and then quickly declines towards higher J values, while the archetypical Seyfert\,2 galaxy NGC\,1068 shows strong CO lines even above J$_{upp}$ = 20. In Mrk\,231, the CO SLED rises up to J$_{upp}$ = 5 and then remains relatively flat for the higher-J transitions, consistent with the presence of a central X-ray source illuminating the circumnuclear region. However, other sources do not seem to follow the expected trends. The starburst galaxy NGC\,253 and the ULIRG NGC\,6240, a mixed source thought to be less dominated by its AGN  than e.g. Mrk\,231 (based on mid-IR diagnostics) exhibit extremely strong high-J lines, suggesting that other excitation sources, like warm PDRs (as in NGC\,253) or strong shocks (as in NGC\,6240) have to be considered as well, i.e.\,that the pure detection of a high-J CO line alone is not an unambiguous signature of an XDR excited by an AGN. In a previous paper (Hailey-Dunsheath et al. 2012), we showed that the well characterized SLED of NGC\,1068 could not be uniquely explained with a mixture of PDR, XDR, and shock models. It was only with additional information that an AGN could be identified as the most likely excitation source of the highly excited components. Our new findings here, drawn from a much larger sample, underline the importance of this caveat.

The implication of these findings is particularly relevant in studies of high-redshift galaxies, where the majority of CO detections are limited to high-J transitions. The findings in this paper demonstrate the difficulty in applying high-J CO diagnostics in a simple manner to identify XDRs/AGNs in dusty high-redshift sources. In particular, if only a single or a few CO lines are observed, without a broad coverage of the entire SLED, these line detections can be misinterpreted.

Furthermore, caution must be exercised in calculations of the total molecular gas mass via the so-called CO-to-H$_2$ conversion factor, $\alpha_{CO}$ = $M_{H_2}$/$L_{CO(1-0)}$, (e.g. expressed in units of M$_{\odot}$/(K kms$^{-1}$pc$^2$)). Since at high redshifts, CO detections are limited to J $>$ 3 lines, many studies are forced to first convert an observed mid-J CO line intensity to a J = 1$\rightarrow$ 0 intensity before applying $\alpha_{CO}$ to arrive at an H$_2$ gas mass. Usually, line ratios between a mid-J or high-J line and the J = 1$\rightarrow$0 line are estimated by comparing the source to a similar galaxy and assuming the typical excitation of the galaxy under consideration \citep{2013ARA&A..51..105C}. However, the diversity in our observed SLEDs (Figure~\ref{fig:CO_SEDs})
suggests that this approach may be problematic; extrapolating from mid or high-J lines to J = 1$\rightarrow$0 using templates based on galaxies with similar inferred physical properties introduces a degree of uncertainty since the high-J to CO(1-0) line ratios can vary by an order of magnitude for otherwise similar sources.

In order to further explore the usefulness of high-J CO lines as a tool for characterizing dusty objects at high redshifts, we construct a ratio-ratio diagram employing just three CO lines: J = 1$\rightarrow$0, 6$\rightarrow$5, and 18$\rightarrow$17. The CO(1-0) transition is traditionally used to trace the total molecular gas mass, while the CO(6-5) line traces a warm and dense component that is generally associated with star formation, and is typically one of the brightest transitions in infrared-bright starburst galaxies (see Figure~\ref{fig:CO_SEDs}). CO(18-17) is a very high-J line that may offer a large leverage in distinguishing starburst-dominated from AGN-dominated excitation. Figure~\ref{fig:ratio_ratio} shows the resulting plot of the CO(18-17)/CO(1-0) versus CO(18-17)/CO(6-5) ratios in cases where flux values (or upper limits) for all three rotational lines are available. Since Figure~\ref{fig:ratio_ratio} is meant to serve as an observational tool, we plot ratios of the observed (unscaled) flux values, as opposed to the line fluxes plotted in Figure~\ref{fig:CO_SEDs} and tabulated in Table~\ref{table:COlines}, which have been scaled to the PACS 9.4" beam size when necessary. (A scaled version of Figure~\ref{fig:ratio_ratio} demonstrates the same qualitative trends discussed below.)

We find that the CO(18-17) line is weak with respect to both CO(1-0) and CO(6-5) in our starburst galaxies (NGC\,253, M\,83, M\,82), and strong in the prototypical Seyfert NGC\,1068 and in the ULIRGs of our sample, which are generally located in a region above the starbursts, in the upper right corner where objects have significant amounts of hot gas. A quantitative treatment of this trend will be presented in section 4.

We also searched for potential trends between the ULIRG positions in this diagram and their AGN properties. We collected the AGN luminosities and AGN fractions, i.e.\,the relative contribution of the AGN (w.r.t. the starburst) to the combined bolometric light in these objects (taken from \citet{2009ApJS..182..628V,2013ApJ...776...27V} and listed in Table~\ref{table:RBGS}), but did not find a clear trend with either of these two parameters.

\subsection{LVG Radiative Transfer Model}
To quantitatively analyze the CO SLEDs of the sources in our sample, we assemble the lower-J line intensities from the literature (see footnotes in Table~\ref{table:COlines}) and employ a large velocity gradient (LVG) model in which the excitation and opacity of the CO lines are determined by the gas density ($n_{H_2}$), kinetic temperature ($T_{kin}$), and the CO-to-H$_2$ abundance per velocity gradient ($\chi_{CO}/(dv/dr)$). We use the escape probability formalism derived for a spherical cloud undergoing uniform collapse, with $\beta$ = (1-$e^{-\tau}$)/$\tau$ \citep{1970MNRAS.149..111C,1974ApJ...189..441G}. Each source is assumed to consist of a large number of these unresolved clouds, such that the absolute line intensities scale with the beam-averaged CO column density, $N_{CO}$. In the following analysis, we assume a canonical value of $\chi_{CO}$ = 10$^{-4}$, motivated by abundance measurements in Galactic molecular clouds \citep{1982ApJ...262..590F}. We assume collisional excitation by H$_2$ assuming an H$_2$ ortho/para ratio of 3.

To carry out the computations, we use the Mark \& Sternberg LVG radiative transfer code described in \cite{2012A&A...537A.133D}, with CO-H$_2$ collisional coefficients taken from \citet{2010ApJ...718.1062Y} and energy levels, line frequencies and Einstein A coefficients taken from the Cologne Database for Molecular Spectroscopy (CDMS). For a given set of parameters, \{$n_{H_2}$, $T_{kin}$, $dv/dr$, $N_{CO}$\}, the code computes the intensities of molecular lines by iteratively solving the equations of statistical equilibrium for the level populations using the escape probability formalism. We calculate a three-dimensional grid of expected CO SLEDs, varying $n_{H_2}$ (10$^{2.4}$-10$^{8.2}$ cm$^{-3}$), $T_{kin}$ (30-2500 K), and $dv/dr$ (0.1-1000 km s$^{-1}$pc$^{-1}$) over a large volume of parameter space. While these parameters determine the shape of the resulting SLED, the line intensity magnitudes are set by the beam-averaged CO column density, $N_{CO}$, which is tweaked to match the observed fluxes and beam sizes.
Thus the intensity of each CO line is given by the expression
\begin{equation}
I_{\rm line} = \frac{h\nu}{4\pi}  x_u A \beta(\tau) N_{\rm CO}
\end{equation}
where $x_u$ is the population fraction in the upper level of the transition,
$h\nu$ is the transition energy,
$A$ is the Einstein radiative coefficient, $\beta(\tau)$ is the escape
probability for a line optical depth $\tau$, and $N_{\rm CO}$ is the
beamed averaged column density.
The molecular gas mass in the beam (for a spherical geometry) is then given by
\begin{equation}
M_{H_2} = \pi R^2 \mu\, m_{H_2} \frac{N_{CO}}{\chi_{CO}}
\end{equation}
where $\mu$ = 1.36 takes into account the helium contribution to the molecular weight and $R$ = $D_A\theta/2$ is the effective radius of the beam, with $D_A$ being the angular diameter distance to the source and $\theta$ the beam size of the line observations.

The LVG-modeling technique thus offers an opportunity to derive the CO-to-H$_2$ conversion factor in an extragalactic source in an independent way, using the gas mass estimate from the source's best-fit LVG model \citep{2013MNRAS.435.2407M}. As mentioned above, the molecular hydrogen gas mass is often obtained from the CO(1-0) line luminosity by adopting a mass-to-luminosity conversion factor, $\alpha_{CO}$. The standard Galactic value, calibrated using several independent methods \citep{1978ApJS...37..407D,1986A&A...154...25B,1987ApJ...319..730S,1988A&A...207....1S}, is $\alpha_{CO} \sim$ 4 - 5 M$_{\odot}$/(K kms$^{-1}$pc$^2$), while subsequent studies of CO emission in ULIRGs found a significantly smaller ratio, $\alpha_{CO} \sim$ 0.8 - 1.0 M$_{\odot}$/(K kms$^{-1}$pc$^2$) \citep{1998ApJ...507..615D}. These values have been adopted by many \citep{1988ApJ...324L..55S, 1990ApJ...362..473T, 1991ApJ...368..112W,2012Natur.486..233W} to convert CO J=1-0 line observations to total molecular gas masses, and the question of what value to apply for what type of galaxy has been a controversial issue for decades. Often, consideration is not given to the dependence of $\alpha_{CO}$ on the average molecular gas conditions (metallicity, temperature, density) in the sources to which they are being applied.

Detailed studies of the conversion factor based on CO SLED modeling of the objects in our sample are beyond the scope of this paper because of the significant uncertainties associated with our simple models. We do estimate, however, the conversion factors derived using the  L$_{CO(1-0)}$ and M$_{H_2}$ values from both the single-component and two-component LVG models for each of the sources in our sample (where the total gas mass, M$_{H_2,total}$ = M$_{H_2,cool}$ + M$_{H_2,warm}$, is used in the two-component case). We note that the sole use of   $^{12}$CO lines to derive the CO-to-H$_2$ conversion factor via LVG modeling raises its own concerns. As discussed in \citet{2013ARA&A..51..207B}, this technique is only sensitive to regions where CO is bright and may therefore miss any component of ``CO-faint" H$_2$, unless combined with observations of other lines, i.e. [C{\scriptsize II}], that trace these ``CO-faint" molecular regions.

\subsection{Fitting Procedure}
To determine the best-fit set of parameters that characterize the CO SLED of each source, we compare the modeled SLEDs to the observed CO line intensities and generate a likelihood distribution for each of the parameters following a Bayesian formalism outlined in \citet{2003ApJ...587..171W} and \citet{2012ApJ...753...70K}. The Bayesian likelihood of the model parameters, $\textbf{\emph{p}}$, given the line measurements, $\textbf{\emph{x}}$, is
\begin{equation}
P(\textbf{\emph{p}}|\textbf{\emph{x}}) = \frac{P(\emph{\textbf{p}})P(\textbf{\emph{x}} | \textbf{\emph{p}})}{\int d\textbf{\emph{p}}P(\textbf{\emph{p}})P(\textbf{\emph{x}}|\textbf{\emph{p}})}
\end{equation}
where $P(\textbf{p})$ is the prior probability of the model parameters and $P(\textbf{\emph{x}}|\textbf{\emph{p}})$ is the probability of obtaining the observed data set given that the source follows the model characterized by $\textbf{\emph{p}}$. Assuming that the measured line strengths have Gaussian-distributed random errors, $P(\textbf{\emph{x}}|\textbf{\emph{p}})$ is the product of Gaussian distributions in each observation,
\begin{equation}
P(\textbf{\emph{x}}|\textbf{\emph{p}}) = \prod_i \frac{1}{\sqrt{2\pi\sigma_i^2}}\exp{\left[-\frac{(x_i-I_i(p))^2}{2\sigma_i^2}\right]}
\end{equation}
where $\sigma_i$ is the standard deviation of the observational measurement for transition $i$ and $I_i\textbf{(\emph{p})}$ is the predicted line intensity for that transition and model. The likelihood distribution of any one parameter is thus the integral of $P(\textbf{\emph{p}}|\textbf{\emph{x}})$ over all the other parameters. In our analysis, we use a binary prior probability, choosing priors that are flat in the logarithm of each parameter, and that go to zero for any model that predicts a molecular gas mass which exceeds the source's dynamical mass (in cases where $M_{dyn}$ is known through other means; see footnote in Table ~\ref{table:LVGparameters}).

We first run a calculation with a single-component model, assuming all the CO lines are emitted by a region characterized by a single kinetic temperature, H$_2$ number density, velocity gradient, and column density. It is instructive to examine how far a single gas component can go in reproducing the entire observed CO SLEDs before a given fit becomes inadequate and a second component must be introduced. However, given the fact that many of the sources considered in this paper have had their low- to mid-J emission lines analyzed in previous studies (see Table~\ref{table:prev_LVG}) using two-component LVG models, we would like to explore how these ``best-fit" multi-component models are modified by the inclusion of the more recent PACS high-J line observations. We therefore divide the line fluxes into two components, one cooler and one warmer (with some mid-J lines receiving significant contributions from both) and follow an iterative procedure, looking for the best solution of one component at a time and subtracting it from the other. This two-step approach thus results in a six parameter model, with three parameters characterizing the shape of the SLED in each respective component. It is therefore not applied to sources which have fewer than three line intensity measurements in each component, i.e. M\,83, since in such cases, the problem is underdetermined, i.e. there are more parameters than data points. The detected transitions from NGC 1068 are also not modeled as arising from two different components in this paper since the full SLED has been thoroughly analyzed and fitted with radiative transfer models in previous papers \citep{2012ApJ...755...57H,2012ApJ...758..108S}.

\section{LVG Results \& Discussion}
The LVG-modeled SLEDs for the single and two-component fits are shown in the left and right-hand panels of Figure~\ref{fig:LVGfit} respectively. Although the single-component LVG models provide only a crude approximation to the complicated mixture of physical conditions that characterize the emitting source, they nonetheless yield surprisingly good fits to the observed SLEDs in most cases. The constraints provided by the PACS high-J CO line detections drive the ``best-fit" kinetic temperatures in these single component LVG models to high values, both relative to their two-component counterparts and relative to CO SLED fits obtained in earlier LVG analyses, when measurements of these high-level excitations were not yet available. Furthermore, demanding that a single set of LVG model parameters reproduce the full CO SLED often results in under-predicted low-J CO transitions. Introducing a separate low-excitation component mitigates both of these issues, accounting for the excess CO(1-0) and CO(2-1) line emissions while yielding a two-component LVG model with more moderate estimates of the kinetic temperature. While a true multi-component model will obviously provide more accurate estimates of these molecular gas properties, the fact that in most cases, a single component LVG model is sufficient to faithfully reproduce the full SED, is a noteworthy result. The CO emission line spectra of star-forming galaxies may often be representable by a single LVG SLED as a function of halo mass and molecular gas mass content.

Below, we present estimates of the best-fit parameters characterizing the molecular gas in each source  and compare our results to those obtained in previous studies (Table~\ref{table:prev_LVG}) when applicable. A summary of these LVG results can be found in Table~\ref{table:LVGparameters}, where ``4D Max" refers to the single most probable grid point in the entire multi-dimensional distribution. The sources are listed and discussed in order of increasing distance.
We also present estimates of the CO-to-H$_2$ conversion factors derived from our best-fit LVG models (Table~\ref{table:alpha}). For the ULIRGs in our template sample, and with the two-component approach, we derive values in the canonical range 0.4 - 0.5 M$_{\odot}$/(K kms$^{-1}$pc$^2$). The lowest factor is found for M\,82 ($\sim$0.2 M$_{\odot}$/(K kms$^{-1}$pc$^2$), while for the other objects it varies between $\sim$0.4 and $\sim$14 M$_{\odot}$/(K kms$^{-1}$pc$^2$). These are all plausible values, keeping in mind that these models refer to the central regions of these starburst, AGN and merger templates. Drawing more firm conclusions from these data sets would require more detailed modeling and a careful analysis of the involved assumptions and error bars. We will address this in future single source studies.

Our LVG results are also applied to further analyze the ratio-ratio diagram (Figure ~\ref{fig:ratio_ratio}) presented in Section 3.1. As previously mentioned, the ULIRGs in our sample tend to be located in the upper right corner of this plot, where objects have significant amounts of hot gas. To obtain a quantitative measure of this trend, we divide 10 of the template objects in Figure~\ref{fig:ratio_ratio} into two  bins by grouping the objects into the lower left and upper right quadrant:
5 objects with CO(18-17)/CO(1-0) $<$ 10 and CO(18-17)/CO(6-5) $<$ 0.2 (lower left quadrant, this group contains NGC\,253, M\,82, NGC\,4945, Circinus, and Arp\,220) and 5 objects with CO(18-17)/CO(1-0) $>$ 10 and CO(18-17)/CO(6-5) $>$ 0.2 (upper right quadrant, this group contains NGC\,1068,  IC\,694, NGC\,3690, NGC\,6240, and Mrk\,231). Using a two-component LVG fit (discussed in sections 3.2 and 3.3), we find that the average fractions of highly-excited warm molecular gas (relative to the total gas mass) for the objects in the lower left  and upper right quadrants are $\sim$11\% and 22\% respectively (with standard deviations of 21\%).
These fractions (found in Table~\ref{table:alpha}) are calculated from the mass estimates given in Table~\ref{table:LVGparameters}, and from \citet{2012ApJ...755...57H} in the case of NGC\,1068.  (Note: M\,83 is not included since we do not have a two-component LVG fit, and consequently, an estimate of the highly-excited gas mass component for this source).

Given the uncertainties in estimating the gas masses and the small size of this subsample, this difference between the percentage of dense and warm gas estimated for each quadrant is not statistically significant. However, the data we have collected so far demonstrate the abundance of hot molecular gas in ULIRGs, both with respect to the warm phase that produces the bulk of the CO luminosity, and with respect to the total molecular gas mass. Detailed single source studies with careful modeling of PDR, XDR and shock components are needed to better understand the mechanisms that determine the large range of CO SLED shapes and the plausible excitation mechanisms.

\subsection{NGC\,253}
NGC\,253 is among the nearest and best-studied starburst galaxies. \citet{2008ApJ...689L.109H} employed a multiline LVG model to analyze NGC\,253's CO SLED, introducing a low-excitation component to model the J=2$\rightarrow$1 and J=1$\rightarrow$0 emission while using the remaining higher transitions (J$_{upp}$ $\leq$ 7), including a $^{13}$CO(6-5) detection with ZEUS, to constrain the high-excitation component. Requiring that  $T_{kin} \leq$ 200 K and restricting the velocity gradient to $dv/dr \sim$ 7 - 40 km s$^{-1}$pc$^{-1}$, they found a low-excitation component characterized by $T_{kin} \leq$ 40 K and $n_{H_2}$ = 10$^{2.4}$ - 10$^{3.0}$ and a high-excitation component with kinetic temperature in the range 80-200 K and $n_{H_2}$ = 10$^{3.8}$ - 10$^{4.1}$, with a velocity gradient of 20 km s$^{-1}$pc$^{-1}$ in both cases.

With the additional CO transition lines observed by SPIRE, \citet{2014A&A...564A.126R} modeled the CO SLED as arising from three distinct molecular gas phases where the low-J lines (J$_{upp}$ $<$ 5) originate from regions characterized by temperatures $T_{kin}$ = 60 K and $T_{kin}$ = 40 K, and number densities n$_{H2}$ = 10$^{3.5}$ cm$^{-3}$  and n$_{H2}$ = 10$^{4.5}$ cm$^{-3}$ respectively, while the mid-to-high-J lines (5 $\leq$ J$_{upp}$ $\leq$ 13) are emitted by a region with a kinetic temperature 110 K and number density of 10$^{5.5}$ cm$^{-3}$.

In our analysis, we limit ourselves to two-phase molecular gas and model our ``high-excitation" component to fit the recent high-J line PACS observations (J=15$\rightarrow$14, 16$\rightarrow$15, and 18 $\rightarrow$17), in addition to  the mid-J lines. We find that the low and high-excitation components are characterized by similar number densities, n$_{H_2}\sim$ 10$^{3.4-3.6}$ cm$^{-3}$, while the temperatures range from 50 K to 1260 K for each respective component. We estimate a total H$_2$ mass of $\sim$ 5$\times$10$^{7}$ M$_{\odot}$, only $\sim$2\% of which is in the warm phase generating the high-J CO line emission.

\subsection{M\,83}
For the nearby, barred starburst galaxy M\,83, \citet{2001AA...371..433I} presented two LVG models to fit the J$_{upper}\leq$ 4 CO emission lines: one with $T_{kin}$ = 30-150 K and $n_{H_2}$ = 10$^{2.7}$-10$^{3.5}$ cm$^{-3}$, and one with $T_{kin}$ = 60-100 K and $n_{H_2}$ = 10$^{3.5}$-10$^{5.0}$ cm$^{-3}$. \citet{2006AA...460..467B}, modeling the J$_{upper} \leq$ 6 CO emission lines as originating from a single region, suggested a fit with $T_{kin}$ = 40 K and $n_{H_2}$ = 10$^{5.8}$ cm$^{-3}$.

With only two high-J line measurements for this source, CO(15-14) and CO(16-15), we do not fit a two-component model in this case due to the underdetermined nature of the problem. However, we find that the observed CO SLED for M\,83 is reasonably well-fitted by a single-component model with a kinetic temperature of $\sim$ 500 K and a H$_2$ number density of 10$^{2.8}$ cm$^{-3}$. The molecular mass traced by the CO emission within the central 325 pc of the source ($\theta \sim$ 21'') in this best-fit model is 1.3$\times$10$^7$ M$_{\odot}$, close to the total gas mass estimate of 3$\times$10$^7$  M$_{\odot}$ derived in \citet{2001AA...371..433I}.

\subsection{M\,82}
Due to its proximity, M\,82 is an extensively studied starburst galaxy. \citet{2010A&A...518L..37P} fit the $^{12}$CO emission spectrum from J=4$\rightarrow$3 to J=13$\rightarrow$12 using a LVG model with $T_{kin}$ = 545 K, $n_{H_2}$ = 10$^{3.7}$ cm$^{-3}$, and $dv/dr$ = 35 km s$^{-1}$pc$^{-1}$. \citet{2012ApJ...753...70K} proposed a two-component model where the cool molecular gas (traced by those lines below J=4$\rightarrow$3) was characterized by a kinetic temperature of 63 K and a H$_2$ number density of 10$^{3.4}$ cm$^{-3}$, while the high-J CO lines traced a very warm gas component with $T_{kin}$ = 447 K and $n_{H_2}$ = 10$^{4.1}$ cm$^{-3}$.

With the inclusion of the additional constraints on the high-excitation component provided by the PACS lines CO(15-14), CO(16-15), and CO(18-17), we find that our fitted parameters compare very well with the two-component model set forth in \citet{2012ApJ...753...70K}. Our LVG analysis yields a warm component with $T_{kin}$ = 500 K and $n_{H_2}$ = 10$^{3.4}$ cm$^{-3}$, and a cool component characterized by a kinetic temperature of 80 K, a H$_2$ number density of 10$^{3.2}$ cm$^{-3}$, and a velocity gradient of $\sim$ 200 km\,s$^{-1}$pc$^{-1}$. This high velocity gradient in the low-excitation component is suspected to be the culprit behind the low CO-to-H$_2$ conversion factor in M\,82, $\alpha_{CO} \sim$ 0.25 M$_{\odot}$/(K km$^{-1}$pc$^2$). Since the optical depth of the J = 1$\rightarrow$0 line is reduced by a large $dv/dr$, less molecular mass is required to produce the observed flux, leading to a reduced estimate of M$_{H_2,tot}$ and consequently, of  $\alpha_{CO}$.
The total H$_2$ mass within the 47"$\times$47" PACS beam is estimated to be $\sim$ 5$\times$10$^7$ M$_\odot$, of the same order as the mass estimates derived in \citet{2003ApJ...587..171W}, \citet{2010A&A...518L..37P}, and \citet{2012ApJ...753...70K}, with 50\% of the mass in the warm component.

\subsection{NGC\,4945 \& Circinus}
At distances of $\sim$ 3.7 and $\sim$ 4 Mpc, NGC\,4945 and Circinus are among the nearest and most infrared-bright spiral galaxies in the sky, with observations of their obscured nuclei classifying them as Seyfert galaxies. In both NGC\,4945 and Circinus, \citet{2008AA...479...75H} used the ratios of the observed integrated intensities of the $^{12}$CO(1-0) to $^{12}$CO(4-3) lines, as well as the $^{13}$CO(1-0) and $^{13}$CO(2-1) transitions to obtain column densities, H$_2$ number densities, and kinetic temperatures for the two sources. Fitting the CO and $^{13}$CO lines, they found a degeneracy in the best-fit parameters in $n_{H_2}$-$T_{kin}$ plane. Their best fit solution for NGC\,4945 was $n_{H_2}$ = 3$\times$10$^4$ cm$^{-3}$ and $T_{kin}$ = 20 K, a solution not significantly better than one with $n_{H_2}$ = 10$^3$ cm$^{-3}$ and $T_{kin}$ = 100 K. \citet{2001A&A...367..457C} found a solution with $n_{H_2}$ = 3$\times$10$^3$ cm$^{-3}$ and $T_{kin}$ = 100 K from $^{12}$CO observations of the three lowest transitions and $^{13}$CO data of the two lowest transitions.

For Circinus, again, \citet{2008AA...479...75H} found a number of solutions that provided consistent CO cooling curves for the low-J transitions. The lowest $\chi^2$ was obtained for $n_{H_2}$ = 10$^4$  cm$^{-3}$ and $T_{kin}$ = 20 K; a second, degenerate, solution was found with $n_{H_2}$ = 10$^3$ cm$^{-3}$ and $T_{kin}$ = 100 K. Their results agreed well with \citet{2001A&A...367..457C}, who found a best-fit solution with $T_{kin}$ = 50-80 K and $n_{H_2}$ = 2$\times$10$^3$ cm$^{-3}$ from observations of the three lowest $^{12}$CO  transitions and the two lowest $^{13}$ CO transitions in this source.

Our LVG analysis of the $^{12}$CO(1-0) to  $^{12}$CO(6-5) lines yields low-excitation components with parameters consistent with those obtained by \citet{2008AA...479...75H}. We find best-fit solutions  with a kinetic temperature of 50 K for both sources, and an H$_2$ number density of 10$^{4.8}$ cm$^{-3}$ and 10$^{4.2}$ cm$^{-3}$ in NGC\,4945 and Circinus, respectively. Modeling the high-J CO emission lines as originating from a separate region leads to a warmer, denser component characterized by $T_{kin}$ = 316 K and n$_{H_2}$ = 10$^5$ cm$^{-3}$ in NGC\,4945, and $T_{kin}$ = 500 K and n$_{H_2}$ = 10$^{4.2}$ cm$^{-3}$ in Circinus. We estimate a total H$_2$ gas mass of 4 - 5$\times$10$^7$ M$_{\odot}$ in both sources, of which only $\sim$ 0.5 - 1\% is in the warm phase.

\subsection{NGC\,1068}
NGC\,1068 is one of the brightest and best studied Seyfert 2 galaxies. Its CO SLED, including the PACS high-J lines, has been extensively analyzed by us \citep{2012ApJ...755...57H} and in the works of \citet{2012ApJ...758..108S}; we direct the reader to those papers and to Table~\ref{table:LVGparameters} for a summary of the LVG-modeled gas excitation in this source. We only note that our single-component model yields an H$_2$ gas mass of 10$^{7}$ M$_{\odot}$, twice the total gas mass estimated by \citet{2012ApJ...755...57H}.

\subsection{NGC\,4418}
NGC\,4418 is a peculiar, single nucleus galaxy, with a
LIRG-like luminosity ($\approx 10^{11}\Lsun$) but with other properties similar
to warm ULIRGs, like a high L$_{FIR}$/M$_{H2}$ ratio, an extreme
[C II] deficit (e.g.\citealt{2011ApJ...728L...7G}), and an extremely compact luminosity source
\citep{2003AJ....125.2341E}.

We find that a cool gas phase characterized by $T_{kin}$ = 50 K and n$_{H_2}$ = 1000 cm$^{-3}$ gives rise to the three lowest $^{12}$CO transitions, while the corresponding high-J lines are emitted by a denser, warmer component with $T_{kin}$ = 100 K and n$_{H_2}$ = 10$^{5.6}$ cm$^{-3}$. The gas mass is estimated to be $\sim$10$^{8}$ and 10$^{9}$ M$_{\odot}$ in each of the respective components, indicating that a significant portion of the molecular gas ($\sim$85\%) is in a dense, warm phase.

\subsection{IC\,694 \& NGC\,3690}

The two galaxies IC\,694 and NGC\,3690 form the luminous infrared merger system known as Arp\,299.
The nuclear region of IC\,694 shows the typical mid-IR characteristics
of ULIRGs (very compact and dust-enshrouded star formation, probably harboring a low-luminosity AGN), while
the nuclear region of NGC\,3690 hosts a Seyfert 2 AGN and is surrounded by regions of star formation.

Strong $^{12}$CO emission has been detected in the nuclei of both IC\,694 and NGC\,3690. \citet{2012ApJ...753...46S} modeled the ratios of the observed integrated intensities of the $^{13}$CO(2-1) transition and the low-J $^{12}$CO lines in these two sources. Their best-fit LVG solutions for IC\,694 and NGC\,3690 had kinetic temperatures ranging from 10 to 500 K and 10 to 1000 K, respectively, and a H$_2$ number density greater than 10$^{2.5}$ cm$^{-3}$ in both cases.

For IC\,694, we find that the low and high-J lines can be modeled as arising from two distinct components characterized by the same kinetic temperature, 200 K, but different H$_2$ number densities, 10$^{3.4}$ cm$^{-3}$ and 10$^{5.8}$ cm$^{-3}$, where the high-J emission originates from the denser component. The molecular gas mass in each of these components is estimated to be $\sim$ 7$\times$10$^{8}$ and 3$\times$10$^{7}$ M$_{\odot}$, respectively. The total gas mass estimate is consistent with the gas mass estimates of the molecular region modeled after the low-J lines in \citet{2012ApJ...753...46S}, as well as the mass estimates derived in \citet{2014arXiv1407.2055R} using only PDR heating. However, as evident from the LVG-modeled CO SLED for this source (Figure~\ref{fig:LVGfit}), this two-component best-fit model under-predicts the observed mid-J lines (9 $\leq$ J$_{upp}$ $\leq$ 14), suggesting that in the case of IC\,694, at least three distinct components are necessary in fitting the full CO SLED.

The observed CO SLED for NGC\,3690 (including the extended emission region where its' galaxy disk overlaps with that of IC\,694), is well-fit by a two-component LVG model in which the low-J lines arise from a cool gas phase (T$_{kin}$ = 50 K) while the high-J line emission originates from a warmer gas phase (T$_{kin}$ = 250 K). These two components, which both contribute to the mid-J line emission, are characterized by similar number densities (4-6$\times$10$^{3}$ cm$^{-3}$) and have a combined gas mass of nearly 10$^9$ M$_\odot$.

\subsection{Arp\,220}
At a distance of about 77 Mpc, Arp\,220 is one of the nearest and best studied ULIRGs, serving as a template for high-$z$ studies of dusty starbursts. We found that a single-component model with $T_{kin}$ = 630 K and $n_{H_2}$ = 10$^{2.8}$ cm$^{-3}$ reproduces the observed CO SLED quite well, yielding an associated H$_2$ gas mass of $\sim$ 10$^{9}$ M$_{\odot}$.

\citet{2011ApJ...743...94R} found that the low-J transitions trace cold gas with $T_{kin}$ = 50 K and $n_{H_2}$ = 10$^{2.8}$ cm$^{-3}$, while the mid-J to high-J lines trace a warmer, denser component with  $T_{kin}$ = 1350 K and $n_{H_2}$ = 10$^{3.2}$ cm$^{-3}$, yielding a total gas mass of $\sim$ 6$\times$10$^9$ M$_{\odot}$. Our two-component LVG analysis yields identical results for the cool molecular gas region; however, we find that the mid to high-J lines are modeled best as arising from a high-excitation component with $T_{kin}$ = 316 K and $n_{H_2}$ = 10$^{4.4}$ cm$^{-3}$. The difference in these values when compared against the best-fit parameters obtained in \citet{2011ApJ...743...94R} for the high-excitation component may be due to the fact that the high-J CO lines are differentially affected by dust extinction, an effect which \citet{2011ApJ...743...94R} argues and accounts for in his modeling, but which was not taken into account in this analysis. Nonetheless, our two-component LVG model yields a total gas mass M$_{H_2}$ $\sim$ 6$\times$10$^9$ M$_{\odot}$, (of which only 2\% is in the warm phase), consistent with the mass estimates derived in \citet{2011ApJ...743...94R}.

\subsection{NGC\,6240}
NGC\,6240 is a nearby luminous infrared galaxy, with a CO SLED similar in shape to that of Mrk\,231. Several X-ray studies have firmly established the presence
of powerful AGN activity (e.g. \citealt{1998MNRAS.297.1219I,1999A&A...349L..57V}), in fact occurring in both nuclei \citep{2003ApJ...582L..15K}. The system is also known for strong shocked emission in the superwind
flow that is driven by the NGC\,6240 starburst \citep{1993ApJ...405..522V,2003A&A...409..867L}.
Although the CO J=8$\rightarrow$7 transition has the largest line intensity, the J$_{upper}$ $\geq$ 14 transitions make a significant contribution, with the intensity of the individual CO lines only slowly decreasing for higher rotational quantum numbers.

Limiting ourselves to a two-component LVG analysis, we find that the low-J lines can be modeled as arising from a region with $T_{kin}$ = 126 K and $n_{H_2}$ = 10$^{3.4}$ cm$^{-3}$, while the high-J lines are produced by a slightly warmer ($T_{kin}$ = 160 K) and much denser gas phase, with a H$_2$ number density of 10$^{7.4}$ cm$^{-3}$. The total gas mass derived for this simplistic model is 2$\times$10$^{10}$ M$_{\odot}$, with close to 60\% of the gas in the warmer, denser phase. However, as in the case of IC\,694, this two-component LVG model fails to reproduce the observed mid-J lines, suggesting that three distinct components may be necessary to fit the full CO SLED. An alternative approach may be to employ shock models to further analyze the CO excitation in NGC\,6240. This was pointed out in \citet{2013ApJ...762L..16M} when they found, among other things, that even the most optimistic estimate of the AGN X-ray luminosity is not enough to explain the combined H$_2$ and CO observed luminosities.

\subsection{Mrk\,231}
Mrk\,231 is the most luminous of the local ULIRGs
and a type 1, low-ionization broad absorption line (LoBAL) AGN.
Limited to observations of the CO ladder up to the J=6$\rightarrow$5 transition in Mrk\,231, \citet{2007ApJ...668..815P} found that while the LVG solutions derived solely from the three lowest CO transitions converged to $T_{kin}$ = 55 -95 K and $n_{H_2}$ $\sim$ 10$^3$ cm$^{-3}$, the CO(4-3) and CO(6-5) emission lines traced a much denser gas phase with $n_{H_2}$ $\sim$ (1-3)$\times$10$^4$ cm$^{-3}$.

With the additional SPIRE FTS \citep{2010AA...518L..42V} and PACS high-J lines, we find that the CO lines up to J = 11-10 can be produced by a two-component LVG model, with a cool region at $T_{kin}$ = 50 K and $n_{H_2}$ = 10$^{3.8}$ cm$^{-3}$ emitting the low-J lines (J$_{upper}$ $<$ 4) and a warmer, denser region at $T_{kin}$ = 316 K and $n_{H_2}$ = 10$^{4.2}$ cm$^{-3}$ dominating the mid- to high-J line emissions. These models yield a total associated gas mass of $\sim$ 10$^9$ M$_{\odot}$, with 20\% of the gas in the high-excitation component. However, a challenge is presented by the highest CO rotational lines (J$_{upper}$ $\geq$ 12), which are strongly under-produced by our two-component LVG model. These higher-J lines may signal the presence of a third excitation component which could be a high-excitation PDR, XDR, or shocks, as explained in \citet{2010AA...518L..42V}.

\section{Summary}
\begin{figure*}[h!]
\begin{minipage}{1\linewidth}
\vspace{1cm}
\hspace{-1.5cm}\includegraphics[width=550pt,height=350pt]{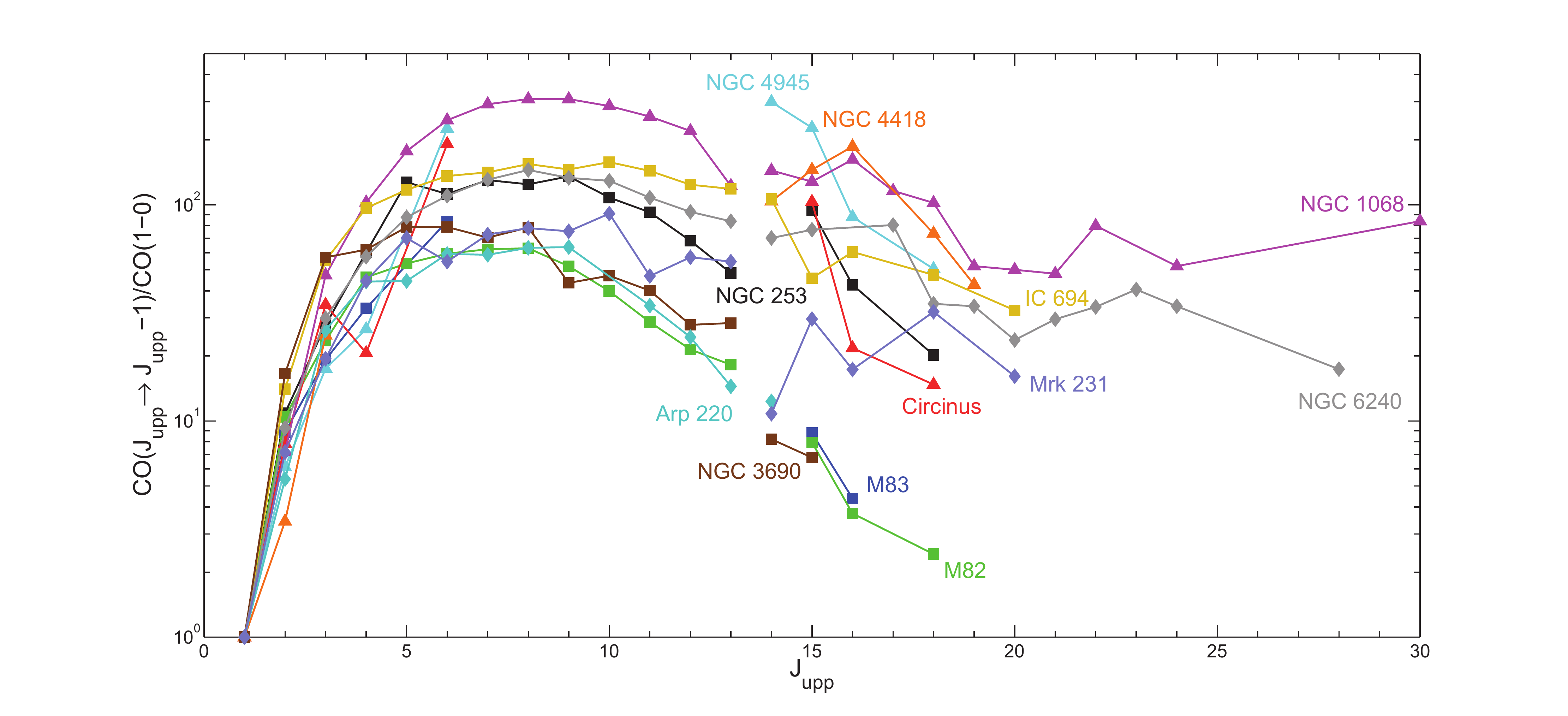}\\
\vspace{-1cm}\caption{Observed CO SLED of sources listed in Table 1. Flux values have been normalized with respect to the CO(1-0) flux measurement using aperture corrections when necessary, as explained in Section \ref{sect:observations}. Starbursts, Seyfert galaxies, and (U)LIRGs are denoted with squares, triangles, and diamonds respectively. The break in the SEDs separates measurements obtained from the literature from the PACS line measurements. }
\label{fig:CO_SEDs}
\end{minipage}
\end{figure*}

\begin{figure*}[h!]
\begin{minipage}{1\linewidth}
\vspace{-3cm}
\includegraphics[width=500pt,height=325pt]{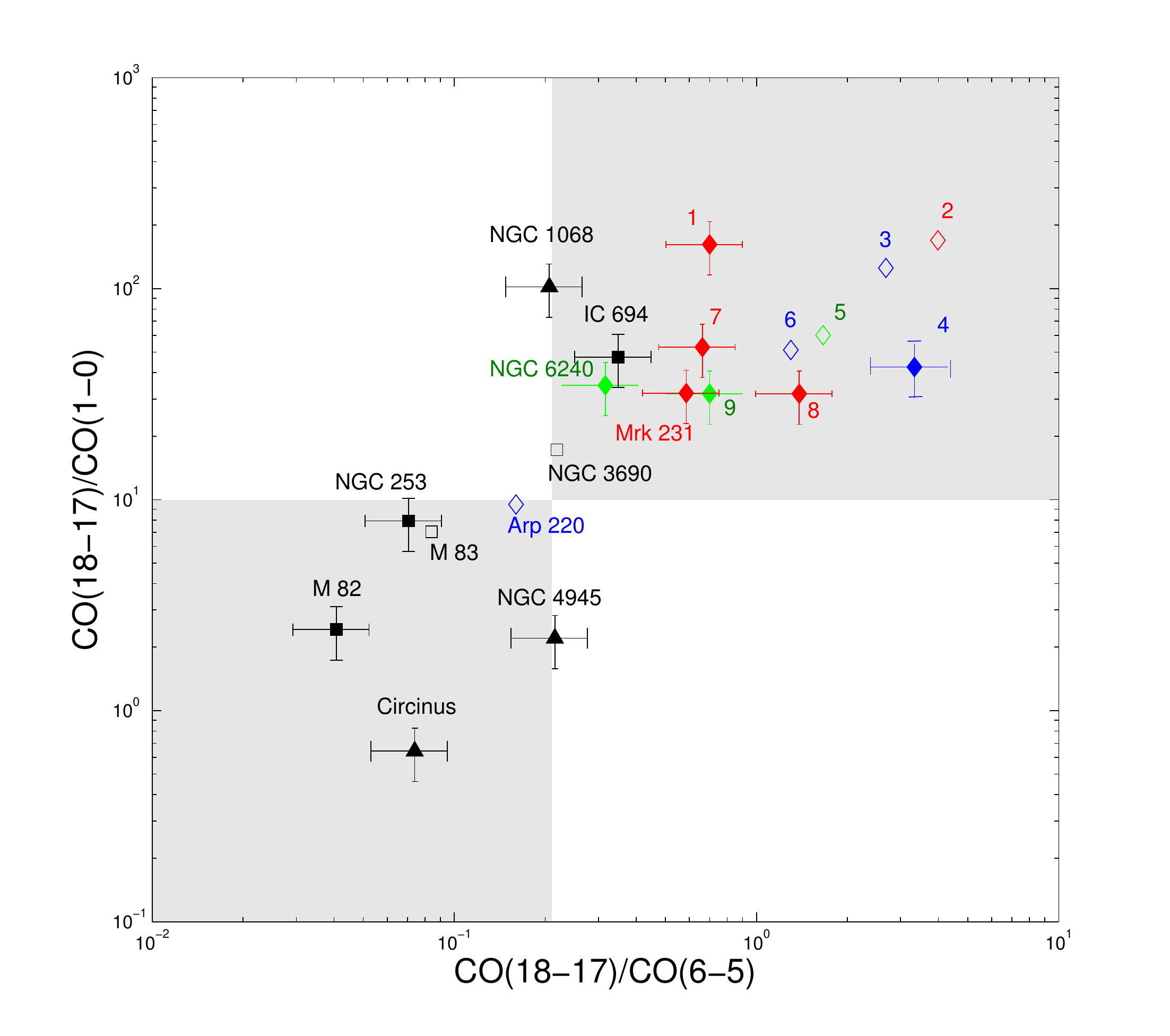}
\vspace{-1cm}\caption{CO(18-17)/CO(1-0) vs. CO(18-17)/CO(6-5) plot. Starbursts, Seyfert galaxies, and (U)LIRGs are denoted with squares, triangles, and diamonds respectively. The ULIRGs are further color-coded according to their AGN fractions (see Table 3), where blue, green, and red indicate AGN fractions between 0-25\%, 25-50\%, and 50-75\% respectively. The RBGS ULIRGs are labeled according to the following: 1. IRAS F08572+3915, 2. IRAS 23128-5919, 3. IRAS F12112+0305, 4. IRAS 17208-0014, 5. IRAS F09320+6134, 6. IRAS F10565+2448, 7. IRAS F05189-2524, 8. IRAS 20551-4250, 9. IRAS 23365+3604. Unfilled markers indicate cases where we only have upper limits on the CO(18-17) and/or CO(6-5) line fluxes; these data points thus represent upper bounds for both ratios in the ratio-ratio plot. }
\label{fig:ratio_ratio}
\end{minipage}
\end{figure*}

We have presented the extragalactic detections of FIR CO rotational line emission within the central 10" of a sample of starburst galaxies, Seyfert galaxies, and (U)LIRGs in the local universe. ($z$ $<$ 0.1). We have augmented our multi-J $^{12}$CO line dataset (J$_{upp} \geq$ 14), detected with Herschel-PACS, with lower-J CO line measurements collected from the literature to yield a well-sampled set of local CO SLEDs (Table~\ref{table:COlines}). Along with providing a necessary benchmark for the usually more sparsely sampled SLEDs obtained for high redshift galaxies, (e.g. \citealt{2007ASPC..375...25W}), these SLEDs demonstrate the uncertainties in relying solely on high-J CO diagnostics to characterize the excitation source of a galaxy. Without a broad coverage of the entire SLED for a given source, a single or few observed CO lines can easily be misinterpreted given the sheer diversity in SLED shapes shown in Figure~\ref{fig:CO_SEDs}.
However, while the detection of a high-J CO line alone is not an unambiguous signature of a particular excitation source, the position of a source on a ratio-ratio diagram may be an indication of the presence, or lack thereof, of a warm, dense molecular gas component.  In the CO(18-17)/CO(1-0) versus CO(18-17)/CO(6-5) plot shown in Figure~\ref{fig:ratio_ratio}, the sources that fall in the upper right corner of the diagram, with CO(18-17)/CO(1-0) $>$ 10 and  CO(18-17)/CO(6-5) $>$ 0.2, consistently have a higher percentage of highly-excited molecular gas than those that fall in the lower left corner, as verified by the LVG-modeling results for these sources.

Another tracer of high density molecular clouds, which is traditionally used in ground-based mm observations, is provided by low HCN lines. These lines trace similar critical densities as the CO lines in the PACS range ($n \sim$ 10$^6$ - 10$^8$ cm$^{-3}$). In fact, a similar trend is found if in the above ratio-ratio plot the x-axis is replaced by the HCN(1-0)/CO(1-0) ratios of our objects as presented in \citet{2004ApJ...606..271G}. However, while both molecules trace dense gas, the energy levels of the upper transitions are different, i.e. the CO traces warmer gas than HCN, and therefore, while HCN lines provide complementary results, they are not 1:1 substitutes for the high-J CO lines. Furthermore, at high redshifts, the CO lines have moved into the (sub-)mm range and are thus better high density tracers than the low-lying HCN transitions which have moved out of that window.

\begin{figure*}[h!]
\vspace{-1.5cm}
\hspace{-1cm}\includegraphics[width=580pt,height=532pt]{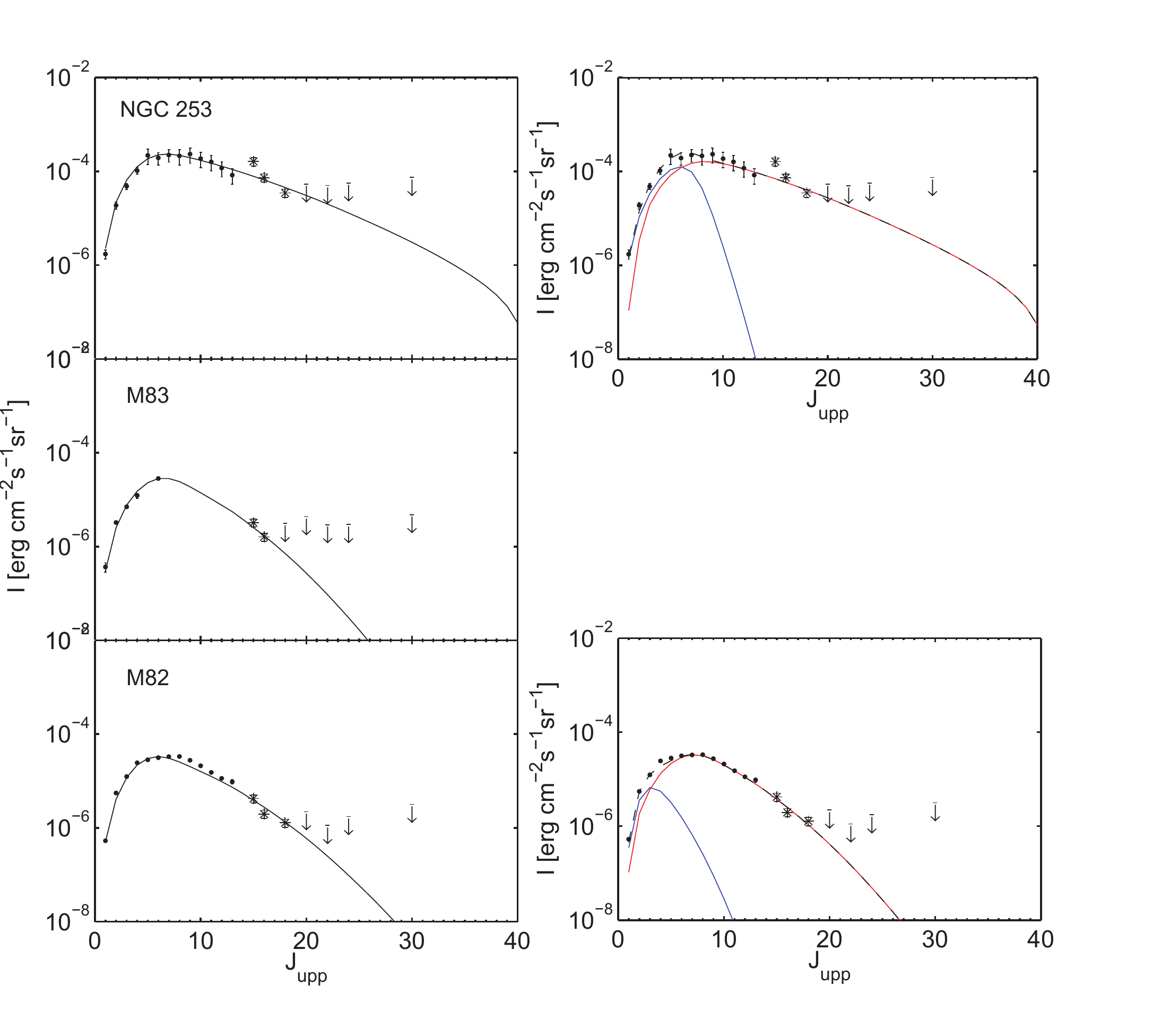}
\vspace{-1.5cm}\caption{SLEDs for $^{12}$CO: Single \emph{(left panel)} and Two-Component \emph{(right panel)} LVG Results. Diamonds represent the low-J line intensities extracted from the data in the references listed in Table 1. Asterisks represent the observed PACS line intensities while arrows signify upper bounds on the line intensity. Best fit SLEDs, corresponding to the "Single/Low/High 4D Max" column in Table 3, are shown with solid/blue/red lines. The dashed line in the two-component fit is the total fitted SLED, i.e., the sum of the two components.}
\label{fig:LVGfit}
\end{figure*}

These results were verified using an LVG radiative transfer modeling technique, which we employed to quantitatively analyze the CO emission from a subset of our detected sources. For each CO SLED, we identify the set of characterizing parameters that best reproduces the observed line intensities. Using both single-component and two-component LVG models to fit the cloud's kinetic temperature, velocity gradient, gas density, and beam-averaged CO column density, we derive the molecular gas mass and the corresponding CO-to-H$_2$ conversion factor for each respective source.  A summary of the best-fit LVG parameters is presented in Table~\ref{table:LVGparameters} and the resulting LVG-modeled SLEDs for the single and two-component fits are shown in the left and right panels of Figure~\ref{fig:LVGfit}, respectively. Remarkably, the CO emission line spectra of star-forming galaxies may often be representable by a single LVG SLED as a function of halo mass and molecular gas mass content. Furthermore, we find that our two-component LVG model results are mostly consistent with the fits obtained in \citet{2014ApJ...795..174K} where the $^{12}$CO SLEDs from J=1$\rightarrow$0 to J=13$\rightarrow$12 are modeled for many of the same sources discussed in this paper.

Estimates of the CO-to-H$_2$ conversion factors derived from our best-fit single and two-component LVG models can be found in Table~\ref{table:alpha}. The relatively low $\alpha_{CO}$ values we find (compared to normal, star-forming galaxies like the Milky Way), are consistent with previous measures of the conversion factor for some of the individual sources as well as the general finding of low $\alpha_{CO}$'s in the center of bright galaxies  \citep{1992A&A...265..447W,1996ApJ...457..678B,1997ApJ...484..702S,1999AJ....117.2632B,1999ApJ...516..114P,2009A&A...506..689Ia,2009A&A...493..525Ib,2012ApJ...753...46S}. Conditions such as low metallicity, high gas temperature, large velocity dispersion, and higher gas density in an extended warm phase outside the GMCs (typical in ULIRGs) all drive $\alpha_{CO}$ to lower values \citep{2013ARA&A..51..207B}. While the high velocity gradient is the primary culprit behind the reduced value of $\alpha_{CO}$ in M\,82 ($\sim$ 0.25 M$_{\odot}$/(K kms$^{-1}$pc$^2$), the bulk of our results are due to the higher temperatures and warm-phase gas densities found in our sample targets (which never approach the low-metallicity regime.) However, given our modest sample size and the small degree of variance we find in the warm gas mass fractions, it becomes difficult to identify any specific trend between these physical conditions and the conversion factor.

We also systematically find higher $\alpha_{CO}$ values for the two-component models which provide separate fits for the low-and high-J lines, than for the single component models where all CO line emissions trace a single region. This trend is consistent with the existence of a dichotomy, most prominent in (U)LIRGS, pointed out in \citet{2012ApJ...751...10Pb}. In their analysis of a CO line survey of (U)LIRGs, \citet{2012ApJ...751...10Pb} find that while one-phase radiative transfer models of the global CO SLEDs yield low $\alpha_{CO}$ values, $< \alpha_{CO} >\,\sim$ 0.6 M$_{\odot}$/(K kms$^{-1}$pc$^2$), in cases where higher-J CO lines allow a separate assessment of gas mass at high densities, near-Galactic $\alpha_{CO} \sim$ (3-6) M$_{\odot}$/(K kms$^{-1}$pc$^2$) values become possible.

In addition to the CO lines presented here, each of the objects of this study also has (SHINING and other) observations of the FIR fine-structure lines. These will provide further constraints on the physical properties of
the photon-dominated regions (PDRs), X-ray dominated regions (XDRs), and shocks in these galaxies, and will
augment the interpretation of the CO emission in future detailed studies.

\acknowledgements
N.M. is supported by the Raymond and Beverly Sackler Tel Aviv University-Harvard Astronomy Program. We thank the DFG for support via German-Israeli Project Cooperation grant STE1869/1-1.GE625/15-1. Basic research in IR astronomy
at NRL is funded by the US ONR; J.F. also acknowledges support from the NHSC. E.G-A is a Research Associate at the Harvard-Smithsonian Center for Astrophysics. A.V. thanks the Leverhulme Trust for a Research Fellowship. S.V. also acknowledges partial support from NASA through Herschel grants 1427277 and 1454738.
PACS has been developed by a consortium of institutes led by MPE (Germany) and including UVIE (Austria); KU Leuven, CSL, IMEC (Belgium); CEA, LAM
(France); MPIA (Germany); INAF-IFSI/OAA/OAP/OAT, LENS, SISSA (Italy); IAC (Spain). This development has been supported by the funding agencies
BMVIT (Austria), ESA-PRODEX (Belgium), CEA/CNES (France), DLR (Germany), ASI/INAF (Italy), and CICYT/MCYT (Spain).

This material is based upon work supported by the National Science Foundation Graduate Research Fellowship under Grant No. DGE1144152. Any opinion, findings, and conclusions or recommendations expressed in this material are those of the authors and do not necessarily reflect the views of the National Science Foundation.

\FloatBarrier

\FloatBarrier

\FloatBarrier
\begin{table}[h!]
\footnotesize
\caption{Observation details of the PACS data}
\vspace{0.2cm}
\begin{minipage}{1\textwidth}
\begin{tabular}{lll}
 \textbf{Name} & \textbf{Type} & \textbf{OBSID}\\
\hline
\hline
NGC\,253 & SB & 1342237601 to -05\\
M\,83 & SB & 1342225788 to -92\\
M\,82 & SB & 1342232254 to -58\\
NGC\,4945 & Sy & 1342212221, 1342247789 to -91\\
Circinus & Sy &1342225144 to -48\\
NGC\,1068 & Sy & 1342191153 and -54, 1342203120 to -30, 1342239374 and -75\\
NGC\,4418 & LIRG & 1342187780, 1342202107 to -16, 1342210830\\
IC\,694 & SB & 1342232602 to -06, 1342232607 and -08\\
NGC\,3690 & SB/AGN & 1342232602 to -046\\
Arp\,220 & ULIRG & 1342191304 to -13\\
NGC\,6240 & ULIRG & 1342240774, 1342216623 and -24\\
Mrk\,231 & ULIRG & 1342186811, 1342207782, 1342253530 to -40\\
Centaurus A & Sy & 1342225986 to -90\\
Antennae & Interacting & 1342234958 to -62\\
NGC\,4039 & Interacting & 1342234953 to -57\\
IRAS 07251-0248 &	ULIRG & 1342207824 and -26\\
IRAS 09022-3615 & ULIRG & 1342209403 and -06\\
IRAS 13120-5453 & ULIRG & 1342214629 and -30\\
IRAS 15250+3609 & ULIRG & 1342213752 and -54\\
IRAS 17208-0014 & ULIRG & 1342229693 and -94\\
IRAS 19542+1110 & ULIRG & 1342208916 and -17\\
IRAS 20551-4250 & ULIRG & 1342208934 and -36\\
IRAS 22491-1808 & ULIRG & 1342211825 and -26\\
IRAS 23128-5919 & ULIRG & 1342210395 and -96\\
IRAS 23365+3604 & ULIRG & 1342212515 and -17\\
IRAS F05189-2524	& ULIRG & 1342219442 and  -45\\
IRAS F08572+3915	& ULIRG & 1342208954 and -55\\
IRAS F09320+6134	& ULIRG & 1342208949 and -50\\
IRAS F10565+2448	& ULIRG & 1342207788 and -90\\
IRAS F12112+0305	& ULIRG & 1342210832 and -33\\
IRAS F13428+5608	& ULIRG & 1342207802 and -03\\
IRAS F14348-1447	& ULIRG & 1342224242 and -44\\
IRAS F14378-3651	& ULIRG & 1342204338 and -39\\
IRAS F19297-0406	& ULIRG & 1342208891 and -93\\
\end{tabular}
\label{table:OBSID}
\end{minipage}
\end{table}

\FloatBarrier
\begin{table}[h!]
\begin{scriptsize}
\vspace{-2cm}\caption{CO Line Observations}
\vspace{0.2cm}
\begin{minipage}{1\textwidth}
\begin{tabular}{lccc}
\textbf{Lines} & & &  \textbf{Flux [10$^{-17}$ W m$^{-2}$]}  \\
\end{tabular}

\begin{tabular}{lcccccc}
   \hline
& NGC\,253%
\tablenotemark{a}
\tablenotetext{a}{\scriptsize \citet{2014A&A...564A.126R, 2008ApJ...689L.109H}}
& M\,83%
  \tablenotemark{b}
\tablenotetext{b}{\scriptsize \citet{2006AA...460..467B, 2001AA...371..433I}}
   & M\,82%
   \tablenotemark{c}
\tablenotetext{c}{\scriptsize \citet{2003ApJ...587..171W, 2012ApJ...753...70K} }
    & NGC\,4945%
   \tablenotemark{d}
\tablenotetext{d}{\scriptsize \citet{2008AA...479...75H}; Weiss (private communication)}
     & Circinus%
       \tablenotemark{d}
        & NGC\,1068%
      \tablenotemark{e}
\tablenotetext{e}{\scriptsize \citet{2011ApJ...736...37K, 2012ApJ...758..108S}} \\
    \hline
 CO(1-0) & 0.4$\pm$0.1 & 0.4$\pm$0.1 & 2.9$\pm$0.2 & 0.19$\pm$0.03 & 0.07$\pm$0.01 & 0.05$\pm$0.001\\
 CO(2-1) & 4.4$\pm$0.7 & 3.8$\pm$0.2 & 29.8$\pm$2.1 & 1.1$\pm$0.2 & 0.52$\pm$0.08 & 0.36$\pm$0.003\\
 CO(3-2) & 11.3$\pm$1.6 & 8.3$\pm$0.7 & 67.0$\pm$5.4 & 3.3$\pm$0.5 & 2.3$\pm$0.3 & 2.4$\pm$0.3 \\
 CO(4-3) & 24.5$\pm$3.7 & 14.3$\pm$1.8 & 131.6$\pm$1.4 & 5.0$\pm$0.7 & 1.4$\pm$0.2 & -- \\
 CO(5-4) & 51.7$\pm$ 18.7 & -- & 152.6$\pm$1.8 & -- & -- & -- \\
 CO(6-5) & 45.6$\pm$13.7 & 36.1$\pm$2.3 & 169.5$\pm$1.0 &  42.2$\pm$8.4 & 12.6$\pm$2.5 & --\\
 CO(7-6) & 52.9$\pm$15.9 & -- & 177.6$\pm$1.7 & -- & --  & --\\
 CO(8-7) & 50.6$\pm$18.2 & -- & 180.0$\pm$2.7 & -- & --  & --\\
 CO(9-8) & 55.1$\pm$19.9 & -- & 148.3$\pm$2.6 & -- & -- & 15.4$\pm$1.6\\
 CO(10-9) & 44.0$\pm$15.9 & -- & 113.7$\pm$1.6 & -- & -- & 14.3$\pm$1.5 \\
 CO(11-10) & 37.6$\pm$13.5 & -- & 81.7$\pm$1.4 & -- & -- & 12.8$\pm$1.3\\
 CO(12-11) & 27.7$\pm$10.0 & -- & 61.0$\pm$1.6 & -- & -- &11.0$\pm$1.1 \\
 CO(13-12) & 19.7$\pm$ 7.1 & -- & 51.9$\pm$5.5 & -- & -- &6.1$\pm$0.7 \\
 \hline
CO(14-13) & -- & -- & -- & 56$\pm$11.2 & --  & 7.2$\pm$2.3\\
CO(15-14) & 38.3$\pm$7.7 & 3.8$\pm$0.8 & 22.7$\pm$4.5 & 42.5$\pm$8.5 & 6.8$\pm$1.4 & 6.4$\pm$2.2\\
   CO(16-15) & 17.4$\pm$3.5 & 1.9$\pm$0.4 & 10.6$\pm$2.1 & 16.4$\pm$3.3 & 1.4$\pm$0.3 & 8.1$\pm$2.5\\
   CO(17-16) & -- & -- & -- & -- & --  & 5.8$\pm$1.8\\
   CO(18-17) & 8.2$\pm$1.6 & $<$ 3.0 & 6.9$\pm$1.4 & 9.5$\pm$1.9 & 1.0$\pm$0.2 & 5.1$\pm$1.6\\
   CO(19-18) & -- & --  & --  & --  & --  & 2.6$\pm$0.9 \\
   CO(20-19) & $<$ 10.6 & $<$ 4.2 & $<$ 9.9 & $<$ 21.0 & $<$ 5.8 & 2.5$\pm$0.9 \\
     CO(21-20) & --  & -- & -- & -- & --  & 2.4$\pm$0.9\\
   CO(22-21) & $<$ 9.8 & $<$ 2.9 & $<$ 5.0 & $<$ 14.9 & $<$ 3.5 & 4.0$\pm$1.4\\
       CO(23-22) & -- & -- & -- & -- & -- & --\\
   CO(24-23) & $<$ 11.1 & $<$ 2.9 & $<$ 7.9 & $<$ 9.1 & $<$ 3.3 & 2.6$\pm$1.0 \\
       CO(25-24) & -- & -- & -- & -- & -- & $<$ 11.2 \\
            CO(26-25) & -- & -- & -- & -- & -- & -- \\
   CO(27-26) & -- & -- & -- & -- & -- & $<$ 5.8\\
   CO(28-27) & -- & -- & -- & -- & -- & $<$ 4.6 \\
   CO(29-28) & -- & -- & -- & -- & -- & $<$ 9.3 \\
   CO(30-29) & $<$ 14.6 & $<$ 4.7 & $<$ 14.2 & $<$ 10.9 & $<$ 4.7 & 4.2$\pm$1.9\\
   \hline
   \hline
\end{tabular}
\vspace{.2cm}
\\With the exception of M\,83, M\,82, and NGC\,3690, the line fluxes recorded in this table represent the fluxes contained in a $\theta$ = 9.4" beam for each source. Flux values have been normalized to this beam size when necessary, using aperture corrections discussed in Section \ref{sect:observations}. In the case of M\,83, M\,82, and NGC\,3690 the fluxes are those contained in $\theta$ = 21", 47", and 19'' beams, respectively.
References for the J$_{upper}$ $<$ 14 line fluxes collected from the literature are given below.
\end{minipage}
\label{table:COlines}
\end{scriptsize}
\end{table}

\FloatBarrier
\begin{table}[h!]
\footnotesize
\vspace{0.2cm}
\vspace{-2cm}
\begin{minipage}{1\textwidth}
\begin{tabular}{lcc}
Table 2 -- \emph{(continued)}: CO Line Observations & &\\
\end{tabular}

\begin{tabular}{lccc}
\textbf{Lines} & & &  \textbf{Flux  [10$^{-17}$ W m$^{-2}$]}   \\
\end{tabular}

\begin{tabular}{lcccccc}
   \hline
        & NGC\,4418%
        \tablenotemark{f}
\tablenotetext{f}{\citet{2013ApJ...764...42S, 2007A&A...475..479A}}
         & IC\,694%
        \tablenotemark{g}
\tablenotetext{g}{\citet{2014arXiv1407.2055R}}
          & NGC\,3690%
          \tablenotemark{g}
           & Arp\,220%
           \tablenotemark{h}
\tablenotetext{h}{\citet{2011ApJ...743...94R, 2009ApJ...692.1432G}}
            & NGC\,6240%
           \tablenotemark{i}
\tablenotetext{i}{\citet{2013ApJ...762L..16M, 2009ApJ...692.1432G} }
             & Mrk\,231%
             \tablenotemark{j}
\tablenotetext{j}{\citet{2010AA...518L..42V,2007ApJ...668..815P} }
             \\
    \hline
C0(1-0) &   0.03$\pm$0.002 & 0.3$\pm$0.1 & 0.09$\pm$0.03 & 0.2$\pm$0.01 & 0.1$\pm$0.01 & 0.03$\pm$0.003\\
CO(2-1) & 0.12$\pm$0.005 & 1.3$\pm$0.4 & 1.5$\pm$0.4 & 0.9$\pm$0.1& 1.1$\pm$0.2 & 0.24$\pm$0.02\\
CO(3-2) &  0.9$\pm$0.1 & 5.1$\pm$1.5 & 5.1$\pm$1.5 & 4.2$\pm$0.5 & 3.6$\pm$0.7 & 0.6$\pm$0.01\\
CO(4-3) & --  & 8.9$\pm$2.7 & 5.6$\pm$1.6 & 7.0$\pm$0.5 & 7.0$\pm$0.6 & 1.5$\pm$0.4\\
CO(5-4) & -- & 10.8$\pm$3.2 & 7.1$\pm$2.1 & 7.0$\pm$0.3 & 10.6$\pm$0.3 & 2.4$\pm$0.5\\
CO(6-5) & -- & 12.5$\pm$3.8 & 7.1$\pm$2.1 & 9.4$\pm$0.2 & 13.3$\pm$0.2 & 1.8$\pm$0.4\\
CO(7-6) & -- & 13.0$\pm$3.9 & 6.3$\pm$1.9 & 9.3$\pm$0.5 & 15.8$\pm$0.2 & 2.4$\pm$0.5 \\
CO(8-7) & -- & 14.2$\pm$4.3 & 7.1$\pm$2.1 & 10.0$\pm$0.6 & 17.5$\pm$0.3 & 2.6$\pm$0.5\\
CO(9-8) & -- & 13.4$\pm$4.0 & 3.9$\pm$1.2 & 10.1$\pm$0.9 & 16.1$\pm$0.3 & 2.5$\pm$0.5\\
CO(10-9)  &-- & 14.5$\pm$4.4 & 4.2$\pm$1.3 & -- & 15.6$\pm$0.3 & 3.1$\pm$0.6 \\
CO(11-10) &  -- & 13.2$\pm$4.0 & 3.6$\pm$1.1 & 5.4$\pm$0.4 & 13.0$\pm$0.3 & 1.6$\pm$0.3\\
CO(12-11) &  -- & 11.4$\pm$3.4 & 2.5$\pm$0.8 & 3.9$\pm$0.3 & 11.2$\pm$0.3 & 1.9$\pm$0.\\
CO(13-12) & -- & 10.9$\pm$3.3 & 2.6$\pm$0.8 & 2.3$\pm$0.5 & 10.1$\pm$0.3 & 1.8$\pm$0.4 \\
\hline
CO(14-13) &  3.6$\pm$0.7 & 9.8$\pm$2.0 & 0.7$\pm$0.1 & 2.0$\pm$0.4  & 8.5$\pm$1.7 & 0.4$\pm$0.1\\
CO(15-14) & 5.0$\pm$1.0 & 4.2$\pm$0.8 & 0.6$\pm$0.1 & -- & 9.3$\pm$1.9 & 1.0$\pm$0.2 \\
CO(16-15) &  6.4$\pm$1.3 & 5.6$\pm$1.1 & $<$ 1.5 & -- & $<$6.1 & 0.6$\pm$0.1 \\
CO(17-16) &  -- & -- & -- & -- & 9.7$\pm$1.9 & --\\
CO(18-17) &  2.5$\pm$0.5 & 4.4$\pm$0.9 & $<$ 1.6 & $<$ 1.5 & 4.2$\pm$0.8 & 1.1$\pm$0.2 \\
CO(19-18) & 1.5$\pm$0.3 & -- & -- & --  & 4.1$\pm$0.8 & -- \\
CO(20-19) &  $<$ 1.3 & 3.0$\pm$0.6 & $<$ 1.4 & -- & 2.9$\pm$0.5 & 0.5$\pm$0.1 \\
CO(21-10) & -- & -- & -- & -- & 3.6$\pm$0.7 & -- \\
CO(22-21) &  -- & $<$ 3.1 & $<$ 1.0 & -- & 4.1$\pm$0.8 & --\\
CO(23-22) & -- & -- & -- & -- & 4.9$\pm$1.0 & -- \\
CO(24-23) &  -- & $<$ 3.5  & $<$1.0 & -- & 4.1$\pm$0.8 & --\\
CO(25-24) &  -- & -- & -- & --  & -- & --\\
     CO(26-25) &  -- & -- & -- & -- & -- & --\\
CO(27-26) &  -- & -- & -- & -- & -- & -- \\
CO(28-27) &  -- & $<$ 2.3 & $<$ 2.9 & -- & 2.1$\pm$0.4  & --\\
CO(29-28) &  -- & -- & -- & -- & -- & --  \\
CO(30-29) &  -- & $<$2.5 & $<$1.3 & -- & $<$ 1.1  & --\\
   \hline
   \hline
\end{tabular}
\end{minipage}
\end{table}

\clearpage

\FloatBarrier
\begin{table}[h!]
\footnotesize
\vspace{-2cm}\caption{Additional PACS CO Line Observations}
\vspace{0.2cm}
\begin{minipage}{1\textwidth}
\begin{tabular}{lccc}
\textbf{Lines} & & &  \textbf{Flux [10$^{-17}$ W m$^{-2}$]}  \\
\end{tabular}

\begin{tabular}{lccc}
   \hline
& Centaurus A & Antennae & NCG\,4039\\
\hline
CO(15-14) & 1.4$\pm$0.3 & $<$ 0.4 & $<$0.5\\
CO(16-15) & $<$ 4.4 & $<$ 1.2 & $<$ 1.8\\
CO(18-17) & $<$ 2.9 & $<$ 1.5 & $<$ 1.4\\
CO(20-19) & $<$ 4.1 & $<$ 1.8 & $<$ 1.6\\
CO(22-21) & $<$ 2.6 & $<$ 1.4 & $<$ 1.3\\
CO(24-23) & $<$2.6 & $<$ 1.6 & $<$ 1.5\\
  \hline
   \hline
\end{tabular}
\end{minipage}
\label{table:addCO}
\end{table}

\FloatBarrier
\begin{table}[h!]
\footnotesize
\caption{The CO J=1-0, 6-5, 18-17, and 20-19 flux%
 \tablenotemark{a}
\tablenotetext{a}{Velocity-integrated line flux densities in central PACS spaxel ($\theta$ = 9.4") in units of 10$^{-19}$ W m$^{-2}$}
measurements for the ULIRG sample}
\vspace{0.2cm}
\begin{minipage}{1\textwidth}
\begin{tabular}{lcccccc}
Name%
\tablenotemark{b}
\tablenotetext{b}{IRAS name}
 & CO(1-0)%
 \tablenotemark{c}
\tablenotetext{c}{\citet{2012ApJÉ758É71Pa,2009AAS...21334906C,1990A&A...236..327M,1997ApJ...478..144S}}
  & CO(6-5)%
   \tablenotemark{d}
   \tablenotetext{d}{20551-4259 and 23128-5919 were detected by us with APEX; rest taken from \citet{2012ApJÉ758É71Pa}}
   &CO(18-17)%
 \tablenotemark{e}
   \tablenotetext{e}{PACS measurements}
 &CO(20-19)%
 \tablenotemark{e}
 & AGN$_{frac}$%
 \tablenotemark{f}
 \tablenotetext{f}{AGN fractions, taken from \citet{2009ApJS..182..628V, 2013ApJ...776...27V}, are included for those ULIRGs that appear in the ratio-ratio plot of Figure 2. The AGN fractions for Arp\,220, NGC\,6240, and Mrk\,231 are 18.5\%, 25.8\%, and 70.9\% respectively. The AGN luminosities for these three sources are 11.5, 11.32, and 12.45 respectively.}
 &Log(L$_{AGN})$%
 \tablenotemark{g}
 \tablenotetext{g}{AGN Luminosity (= AGN$_{frac}$ * L$_{bol}$)}
   \\
 & & & & & [\%] & [\Lsun]\\
\hline
\hline
F08572+3915 & 0.4$\pm$0.06 & 92$\pm$36 & 65$\pm$10 & 29$\pm$11 & 71.6 & 12.07\\
23128-5919 & 1.8 & 75 & $<$ 300 & $<$ 60 & 63.0 & 11.89\\
F12112+0305 & 1.6$\pm$0.2 & 75$\pm$33 & $<$ 200 & $<$100 & 9.5 & 11.36\\
17208-0014 & 6.1$\pm$0.6 & 78$\pm$32 & 259$\pm$77 & $<$ 80 & 10.9 & 11.54\\
F09320+6134 & 2.7$\pm$0.5 & $<$ 96 & $<$ 160 & 83$\pm$21 & 54.9 & 11.59\\
 F10565+2448 &  2.9$\pm$0.3 & 115$\pm$34 & $<$ 150 & $<$ 60 & 16.6 & 11.33\\
F05189-2524 & 1.8$\pm$0.3 &146$\pm$50 & 96$\pm$18 & $<$ 50 & 71.3 & 12.07\\
  20551-4250 & 2.9 & 66 & 92$\pm$29 & 42$\pm$12 & 56.9 & 11.87\\
  23365+3604 & 1.5$\pm$0.2 & 67$\pm$25 & 47$\pm$18 & $<$ 60 & 44.6 & 11.87 \\
07251-0248 & -- & -- & $<$ 90 & $<$ 50 &\\
09022-3615 & -- & -- & $<$ 70 & 31.4$\pm$8.7 &\\
 13120-5453 & -- & -- & 258$\pm$68 & 289$\pm$43 &\\
 15250+3609 & 0.5$\pm$0.1 & -- & 49$\pm$18 & $<$ 60 &\\
19542+1110 & -- & -- & -- & $<$ 100 &\\
 22491-1808 & 1.3$\pm$0.2 & -- & $<$ 60 & 47$\pm$18 &\\
F13428+5608 & 3.1$\pm$0.3 & -- & $<$  200 & $<$80 &\\
 F14348-1447 & 2.1$\pm$0.3 & -- & $<$  100 & $<$100 &\\
 F14378-3651 & 0.9 & -- &  $<$  70 & $<$30 &\\
 F19297-0406  & 1.3 & -- & $<$  80 & $<$80 &\\
\end{tabular}
\end{minipage}
\label{table:RBGS}
\end{table}

\FloatBarrier

\FloatBarrier
\begin{figure}[h!]
\begin{minipage}{1\linewidth}
\vspace{-1.5cm}
\hspace{-1cm}
\includegraphics[width=580pt,height=532pt]{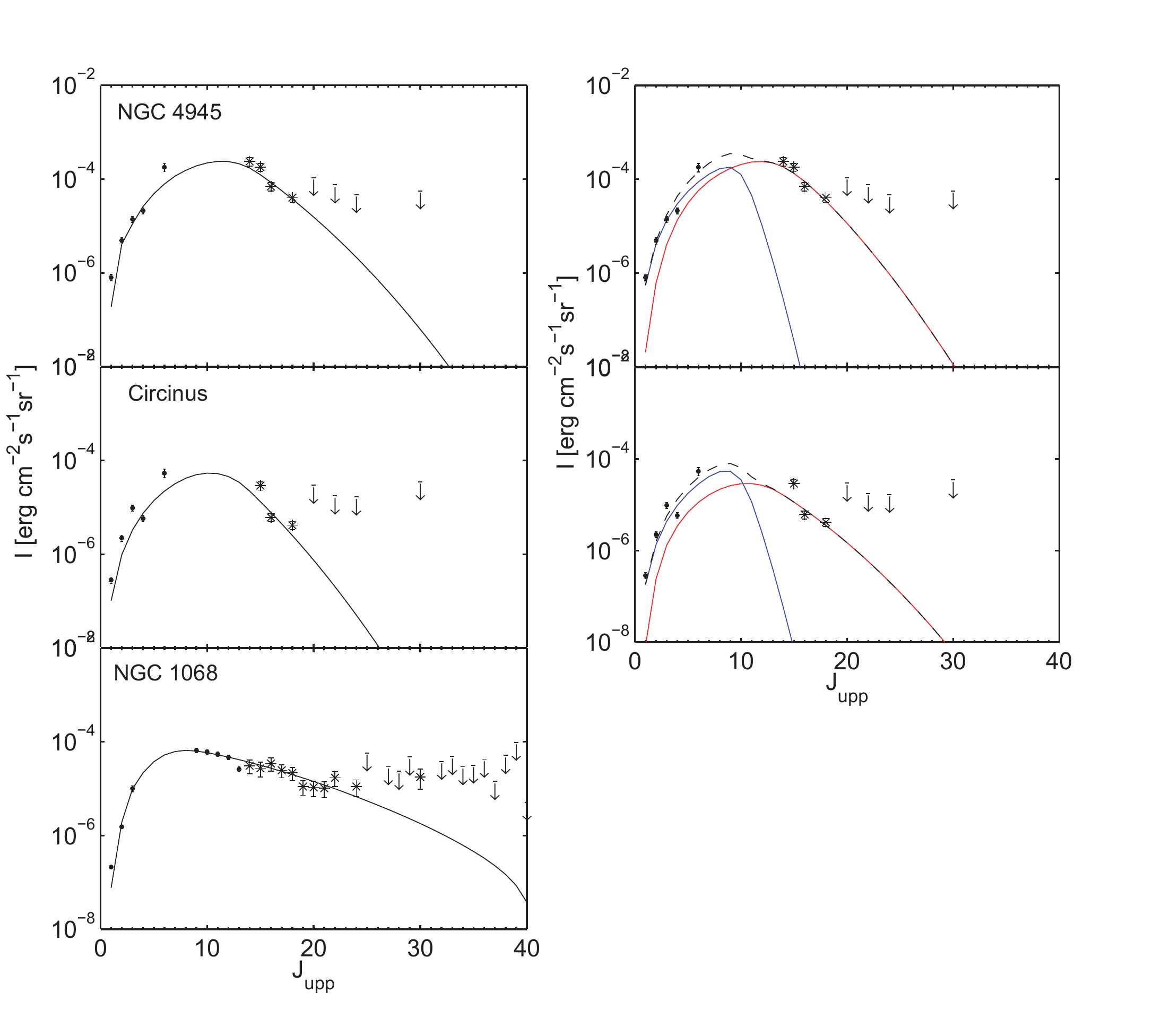}
 Fig.    3.\emph{(continued)}-- SLEDs for $^{12}$CO: Single \emph{(left panel)} and Two-Component \emph{(right panel)} LVG Results.
\end{minipage}
\end{figure}

\FloatBarrier
\begin{figure}[h!]
\begin{minipage}{1\linewidth}
\vspace{-1.5cm}
\hspace{-1cm}
\includegraphics[width=580pt,height=532pt]{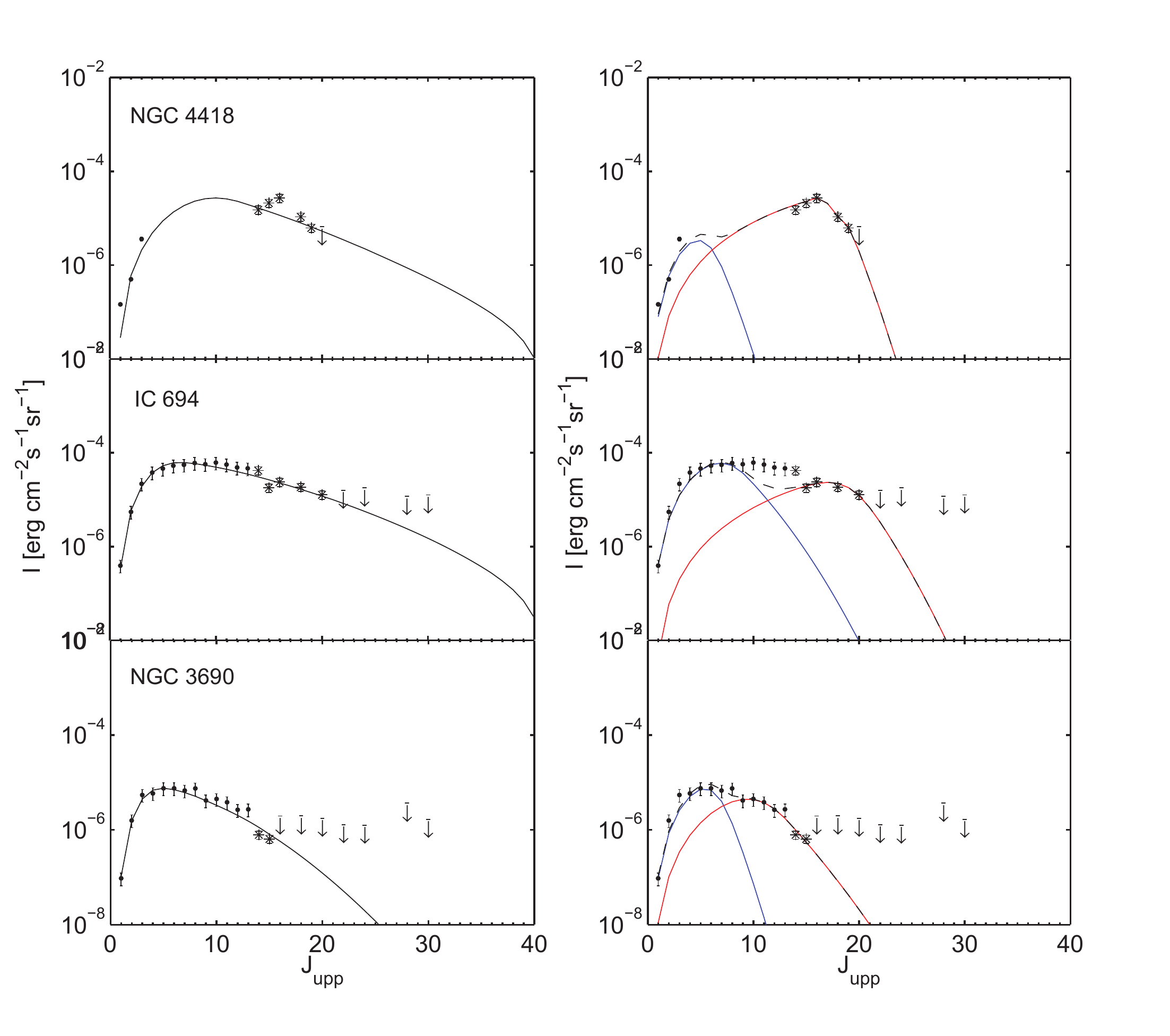}
 Fig.    3.\emph{(continued)}-- SLEDs for $^{12}$CO: Single \emph{(left panel)} and Two-Component \emph{(right panel)} LVG Results.
\end{minipage}
\end{figure}

\FloatBarrier
\begin{figure}[h!]
\begin{minipage}{1\linewidth}
\vspace{-1.5cm}
\hspace{-1cm}
\includegraphics[width=580pt,height=532pt]{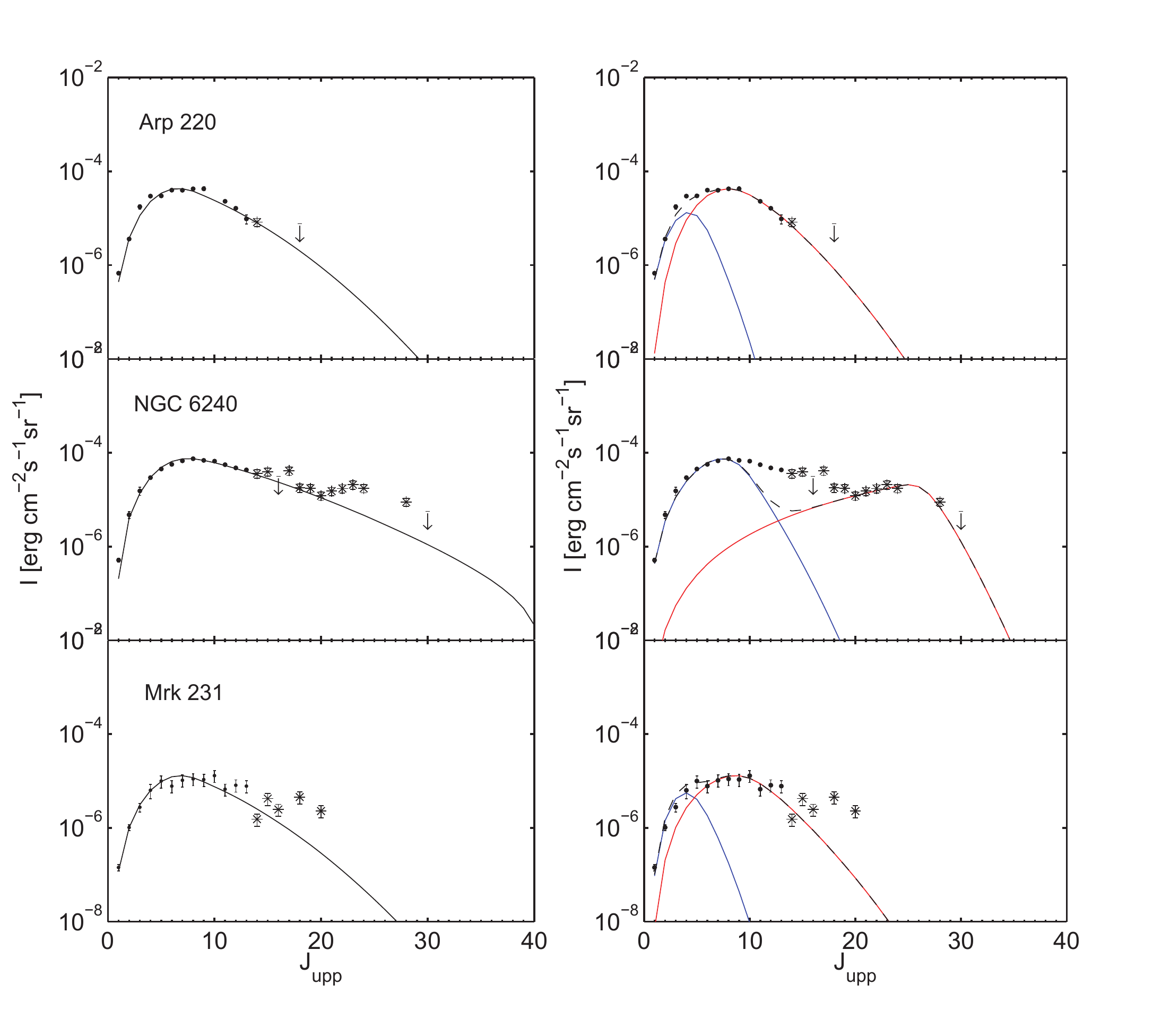}
 Fig.    3.\emph{(continued)}-- SLEDs for $^{12}$CO: Single \emph{(left panel)} and Two-Component \emph{(right panel)} LVG Results.
\end{minipage}
\end{figure}

\FloatBarrier
\begin{table*}[h!]
\footnotesize
\vspace{-2.7cm}
\caption{LVG Model: Best Fit Parameters \& Results}
\vspace{.5cm}
\begin{tabular}{ ccccccc }
& Single & & Low & & High\\
 \hline
 & 4D Max & Range & 4D Max & Range & 4D Max & Range\\
  \hline
  \hline
  \multicolumn{3}{l}{NGC\,253} \\
  \hline
   T [K] & 1260 & 1000 - 1585  & 50 & 30 - 70 & 1260 & 1000 - 1585\\
 n$_{H_2}$ [cm$^{-3}]$ & 10$^{2.6}$ & 10$^{2.4}$ -  10$^{3.2}$ & 10$^{3.4}$ & 10$^{3.2}$ - 10$^{4.0}$ & 10$^{3.6}$ & 10$^{3.2}$ - 10$^{3.8}$ \\
 dv/dr [km s$^{-1}$pc$^{-1}$] & 1 & 0.1 - 10 & 1 & 0.1 - 3 & 32 & 5 - 300\\
 N$_{H_2}$ [cm$^{-2}$] &  10$^{22.6}$ & 10$^{21.9}$ - 10$^{22.9}$ & 10$^{23.1}$ & 10$^{23.0}$ - 10$^{24.9}$ & 10$^{21.5}$ & 10$^{21.2}$ - 10$^{21.9}$ \\
 M$_{H_2}$[M$_{\odot}$]  & 10$^{7.1}$ &  10$^{6.5}$ - 10$^{7.4}$ & 10$^{7.7}$ & 10$^{7.5}$ - 10$^{9.4}$ & 10$^{6.0}$ & 10$^{5.8}$ - 10$^{6.5}$\\
  \hline
   \hline
  \multicolumn{3}{l}{M\,83} \\
  \hline
   T [K] & 500 & 400 - 630 & &  & &   \\
 n$_{H_2}$ [cm$^{-3}$] & 10$^{2.8}$ & 10$^{2.6}$ - 10$^{3.0}$ &  &  &  &\\
 dv/dr [km s$^{-1}$pc$^{-1}$] & 1 & 0.1 - 2 &  & &  &  \\
 N$_{H_2}$ [cm$^{-2}$] &  10$^{21.8}$ & 10$^{21.7}$ - 10$^{22.1}$ &  &  &   &  \\
 M$_{H_2}$ [M$_{\odot}$]  & 10$^{7.1}$ &  10$^{7.0}$ - 10$^{7.5}$ &  &   & &  \\
  \hline
   \hline
   \multicolumn{3}{l}{M\,82} \\
  \hline
 T [K] & 630 & 500 - 794 & 80 & 20 - 400 & 500 & 250 - 500\\
 n$_{H_2}$ [cm$^{-3}]$ & 10$^{2.6}$ & 10$^{2.4}$ -  10$^{2.8}$ & 10$^{3.2}$ & 10$^{2.6}$ -  10$^{3.4}$ & 10$^{3.4}$ & 10$^{3.4}$ -  10$^{4.0}$\\
 dv/dr [km s$^{-1}$pc$^{-1}$] & 1 & 0.1 - 2 & 200 & 20 - 1000 & 8 & 1 - 1000 \\
 N$_{H_2}$ [cm$^{-2}$] &  10$^{21.9}$ & 10$^{21.8}$ - 10$^{22.3}$ &  10$^{21.3}$ & 10$^{20.9}$ - 10$^{22.5}$ &  10$^{21.3}$ & 10$^{20.7}$ - 10$^{21.7}$\\
 M$_{H_2}$ [M$_{\odot}$]  & 10$^{8.0}$ &  10$^{7.9}$ - 10$^{8.4}$ & 10$^{7.4}$ &   10$^{7.0}$ - 10$^{8.6}$ & 10$^{7.4}$ &   10$^{6.8}$ - 10$^{7.8}$\\
  \hline
  \hline
   \multicolumn{3}{l}{NGC\,4945} \\
  \hline
 T [K] & 500 & 400 - 630& 50 & 30 - 60 & 316 & 200 - 630 \\
 n$_{H_2}$ [cm$^{-3}]$ & 10$^{3.6}$ & 10$^{3.4}$ -  10$^{3.8}$ & 10$^{4.8}$ & 10$^{3.8}$ -  10$^{5.0}$ & 10$^{5.0}$ & 10$^{3.8}$ -  10$^{5.4}$\\
 dv/dr [km s$^{-1}$pc$^{-1}$] & 1.0 & 0.1 - 1 & 20 & 0.1 - 25 & 250 & 250 - 1000\\
 N$_{H_2}$ [cm$^{-2}$] &  10$^{22.1}$ & 10$^{22.0}$ - 10$^{22.9}$ &  10$^{22.9}$ & 10$^{22.5}$ - 10$^{23.9}$ &  10$^{21.1}$ &10$^{20.5}$ - 10$^{21.3}$\\
 M$_{H_2}$ [M$_{\odot}$]  & 10$^{6.9}$ &  10$^{6.7}$ - 10$^{7.6}$ & 10$^{7.6}$ &  10$^{7.3}$ - 10$^{8.6}$ & 10$^{5.8}$ &  10$^{5.2}$ - 10$^{6.1}$\\
  \hline
    \hline
   \multicolumn{3}{l}{Circinus} \\
  \hline
 T [K] & 316 & 250 - 500 & 50 & 30 - 60 & 500 & 400 - 1000 \\
 n$_{H_2}$ [cm$^{-3}]$ & 10$^{3.6}$ & 10$^{3.4}$ -  10$^{3.8}$ & 10$^{4.2}$ & 10$^{3.8}$ -  10$^{5.0}$ & 10$^{4.2}$ & 10$^{3.6}$ -  10$^{4.8}$\\
 dv/dr [km s$^{-1}$pc$^{-1}$] & 1 & 0.1 - 2 & 2 & 0.1 - 32 & 25 & 2 - 1000\\
 N$_{H_2}$ [cm$^{-2}$] &  10$^{21.7}$ & 10$^{21.2}$ - 10$^{22.6}$ &  10$^{22.9}$ & 10$^{22.0}$ - 10$^{24.1}$ &  10$^{20.6}$ & 10$^{19.7}$ - 10$^{21.3}$\\
 M$_{H_2}$ [M$_{\odot}$]  & 10$^{6.5}$ &  10$^{5.9}$ - 10$^{7.4}$ & 10$^{7.7}$ &  10$^{6.8}$ - 10$^{8.9}$ & 10$^{5.4}$ &  10$^{4.5}$ - 10$^{6.0}$\\
 \hline
   \hline
   \multicolumn{3}{l}{NGC\,1068} \\
  \hline
 T [K] & 1585 & 1260 - 2000 & & & & \\
 n$_{H_2}$ [cm$^{-3}]$ & 10$^{3.4}$ & 10$^{3.2}$ -  10$^{3.6}$ & & & &\\
 dv/dr [km s$^{-1}$pc$^{-1}$] & 20.0 & 13 - 32 & & & &\\
 N$_{H_2}$ [cm$^{-2}$] &  10$^{21.2}$ & 10$^{20.9}$ - 10$^{21.4}$ & & & &\\
 M$_{H_2}$ [M$_{\odot}$]  & 10$^{7.0}$ &  10$^{6.8}$ - 10$^{7.3}$ & & & &\\

      \end{tabular}
\label{table:LVGparameters}

  \tablenotetext{*}{With the exception M\,83, M\,82, and NGC\,3690, for which the molecular gas masses contained within $\theta \sim$ 21'', 47'', and 19'' beams are given respectively, all other estimates of $M_{H_2}$ correspond to the molecular gas mass in a $\theta$ = 9.4" beam centered on the emission region}
  \tablenotetext{**}{Dynamical masses used in restricting LVG models: M$_{dyn} \sim$ 2$\times$10$^9$ (M\,83, M\,82), 3$\times$10$^9$ (NGC\,4945, Circinus), 9$\times$10$^8$ (NGC\,1068), 3.4$\times$10$^9$ (IC\,694), 1.3$\times$10$^9$ (NGC\,3690), 4$\times$10$^{10}$ (Arp\,220), 6$\times$10$^{10}$ (NGC\,6240), 3.3$\times$10$^{10}$ M$_{\odot}$ (Mrk\,231). (References: \citealt{2001AA...371..433I}, \citealt{2012ApJ...753...70K}, \citealt{2008AA...479...75H}, \citealt{2012ApJ...755...57H}, \citealt{2012ApJ...753...46S}, \citealt{2009ApJ...692.1432G}, \citealt{2007ApJ...668..815P})}

 \end{table*}

\FloatBarrier
  \begin{table*}[h!]
\footnotesize
 \vspace{-2.7cm}
  Table 5 -- \emph{(continued)}: LVG Models: Best Fit Parameters \& Results\\

    \vspace{.2cm}

\begin{tabular}{ ccccccc }
& Single & & Low & & High\\
 \hline
 & 4D Max & Range & 4D Max & Range & 4D Max & Range\\
\hline
\hline
   \multicolumn{3}{l}{NGC\,4418} \\
  \hline
  T [K] & 1260 & 1000 - 1585 & 50 & 20 - 100 & 100 & 63 - 125 \\
 n$_{H_2}$ [cm$^{-3}]$ & 10$^{3.4}$ & 10$^{3.2}$ -  10$^{3.8}$ & 10$^{3.0}$ &10$^{2.6}$ -  10$^{3.4}$ & 10$^{5.6}$ &10$^{5.2}$ -  10$^{5.8}$\\
 dv/dr [km s$^{-1}$pc$^{-1}$] & 3 & 2 - 4 & 1 & 0.1 - 8 & 3 & 0.1 - 5\\
 N$_{H_2}$ [cm$^{-2}$] &  10$^{20.9}$ & 10$^{20.5}$ - 10$^{21.0}$ & 10$^{21.7}$ &10$^{20.9}$ - 10$^{22.8}$ & 10$^{22.4}$ &  10$^{21.8}$ - 10$^{23.6}$\\
 M$_{H_2}$ [M$_{\odot}$]  & 10$^{7.4}$ &  10$^{7.0}$ - 10$^{7.5}$ & 10$^{8.2}$ & 10$^{7.5}$ - 10$^{9.3}$ & 10$^{9.0}$ &  10$^{8.3}$ - 10$^{10.1}$\\
  \hline
    \hline
    \multicolumn{3}{l}{IC\,694} \\
  \hline
 T [K] &1585 & 1260 - 2000 & 200 & 160 - 250 & 200 & 100 - 500 \\
 n$_{H_2}$ [cm$^{-3}]$ & 10$^{3.0}$ & 10$^{2.8}$ -  10$^{3.2}$ & 10$^{3.4}$  & 10$^{3.2}$ -  10$^{3.6}$ & 10$^{5.8}$ & 10$^{3.8}$ -  10$^{5.8}$\\
 dv/dr [km s$^{-1}$pc$^{-1}$] & 13 & 8 - 25 & 3 & 1 - 6 & 126 & 6 - 1000 \\
 N$_{H_2}$ [cm$^{-2}$] &  10$^{21.6}$ & 10$^{21.3}$ - 10$^{21.8}$ & 10$^{22.0}$ & 10$^{21.7}$ - 10$^{22.3}$ &  10$^{20.6}$ & 10$^{19.5}$ - 10$^{21.6}$\\
 M$_{H_2}$ [M$_{\odot}$]  & 10$^{8.4}$ &  10$^{8.1}$ - 10$^{8.7}$ & 10$^{8.8}$ & 10$^{8.6}$ - 10$^{9.2}$   & 10$^{7.5}$ & 10$^{6.3}$ - 10$^{8.5}$\\
  \hline
  \hline
  \multicolumn{3}{l}{NGC\,3690} \\
  \hline
T [K] & 630 & 500 - 800 &50 & 30 - 70 &  250 & 126 - 400\\
 n$_{H_2}$ [cm$^{-3}]$ & 10$^{3.0}$ & 10$^{2.8}$ -  10$^{3.2}$  & 10$^{3.8}$ & 10$^{3.4}$ -  10$^{4.2}$ & 10$^{3.6}$ & 10$^{3.4}$ -  10$^{5.4}$\\
 dv/dr [km s$^{-1}$pc$^{-1}$] & 40 & 16 - 126  &16 & 1 - 50 & 1 & 0.1 -- 100\\
 N$_{H_2}$ [cm$^{-2}$] &  10$^{20.9}$ & 10$^{20.6}$ - 10$^{21.2}$  & 10$^{21.3}$ & 10$^{20.9}$ - 10$^{22.7}$ & 10$^{20.7}$ & 10$^{19.7}$ - 10$^{21.7}$\\
 M$_{H_2}$ [M$_{\odot}$]  & 10$^{8.4}$ &  10$^{8.1}$ - 10$^{8.6}$  & 10$^{8.8}$ &  10$^{8.4}$ - 10$^{10.2}$ & 10$^{8.2}$ &  10$^{7.1}$ - 10$^{9.2}$\\
 \hline
 \hline
\multicolumn{3}{l}{Arp\,220}\\
  \hline
 T [K] & 630 & 500 - 1000 & 50 & 20 - 63 & 316 & 200 - 400\\
 n$_{H_2}$ [cm$^{-3}]$ & 10$^{2.8}$ & 10$^{2.4}$ -  10$^{3.0}$ & 10$^{2.8}$ & 10$^{2.4}$ -  10$^{3.2}$ & 10$^{4.4}$ & 10$^{3.2}$ -  10$^{4.8}$\\
 dv/dr [km s$^{-1}$pc$^{-1}$] & 1 & 0.1 - 2 & 1 & 0.1 - 10 & 32 & 1 - 1000\\
 N$_{H_2}$ [cm$^{-2}$] &  10$^{21.9}$ & 10$^{21.6}$ - 10$^{22.4}$ &  10$^{22.3}$ & 10$^{21.3}$ - 10$^{23.5}$ &  10$^{20.7}$ &  10$^{20.3}$ - 10$^{21.6}$\\
 M$_{H_2}$ [M$_{\odot}$]  & 10$^{9.2}$ &  10$^{9.0}$ - 10$^{9.7}$ & 10$^{9.7}$ &  10$^{8.7}$ - 10$^{10.8}$ & 10$^{8.0}$ &  10$^{7.6}$ - 10$^{8.9}$\\
  \hline
  \hline
\multicolumn{3}{l}{NGC\,6240} \\
  \hline
 T [K] & 1260 & 1000 - 1585 & 126 & 100 - 160 & 160 & 100 - 200\\
 n$_{H_2}$ [cm$^{-3}]$ & 10$^{3.2}$ & 10$^{3.0}$ -  10$^{3.4}$ & 10$^{3.4}$ & 10$^{3.2}$ -  10$^{3.8}$ & 10$^{7.4}$ &  10$^{7.2}$ - 10$^{8.2}$\\
 dv/dr [km s$^{-1}$pc$^{-1}$] &  8 & 4 - 10 & 1 & 0.1 - 2 & 20 & 4 - 25 \\
 N$_{H_2}$ [cm$^{-2}$] &  10$^{21.5}$ & 10$^{21.2}$ - 10$^{21.7}$ &  10$^{22.4}$ & 10$^{22.0}$ - 10$^{22.5}$ &  10$^{22.5}$ & 10$^{22.0}$ - 10$^{24.4}$\\
 M$_{H_2}$%
   \tablenotemark{\dagger} [M$_{\odot}$]  & 10$^{9.1}$ &  10$^{8.8}$ - 10$^{9.3}$ & 10$^{10.0}$ &  10$^{9.6}$ - 10$^{10.1}$ & 10$^{10.1}$ & 10$^{9.6}$ - 10$^{12.0}$\\
  \hline
   \hline
\multicolumn{3}{l}{Mrk\,231} \\
  \hline
 T [K] & 630 & 500 - 794 & 50 & 20 - 63 & 316 & 160 - 500\\
 n$_{H_2}$ [cm$^{-3}]$ & 10$^{2.8}$ & 10$^{2.6}$ -  10$^{3.0}$ & 10$^{3.8}$ & 10$^{3.6}$ -  10$^{4.0}$ & 10$^{4.2}$ & 10$^{3.2}$ -  10$^{4.8}$\\
 dv/dr [km s$^{-1}$pc$^{-1}$] &  1 & 0.1 - 2 & 200 & 160 - 250 & 50 & 2 - 1000\\
 N$_{H_2}$ [cm$^{-2}$] &  10$^{21.4}$ & 10$^{21.3}$ - 10$^{21.6}$ &  10$^{21.0}$ & 10$^{20.8}$ - 10$^{21.5}$ &  10$^{20.4}$ &  10$^{19.7}$ - 10$^{21.5}$\\
 M$_{H_2}$ [M$_{\odot}$]  & 10$^{9.4}$ &  10$^{9.3}$ - 10$^{9.6}$ & 10$^{9.0}$ &  10$^{8.8}$ - 10$^{9.5}$ & 10$^{8.4}$ & 10$^{7.7}$ - 10$^{9.6}$   \\

   \end{tabular}
 \end{table*}

\FloatBarrier
\begin{table}[h!]

\vspace{-1.7cm}\caption{LVG Results from Previous Studies}
\begin{minipage}{1\textwidth}
\footnotesize

\centering
\vspace{.5cm}\begin{tabular}{ccccccc }
\hline
\textbf{Source}  & \textbf{Component} & $\textbf{\emph{T}}$ & \textbf{\emph{n$_{H_2}$}} & $\textbf{\emph{dv/dr}}$ & $\textbf{\emph{N$_{CO}$}}$  & \textbf{\emph{M}}$_{H_2}$\\
& (low/high J) & [K] & [cm$^{-3}$] & [km s$^{-1}$pc$^{-1}$] & [cm$^{-2}$] & [M$_\odot$]\\
\hline
\hline
NGC\,253%
\tablenotemark{a}
\tablenotetext{a}{\citet{2008ApJ...689L.109H}; $\theta \sim$ 11''}
           & low & $\leq$ 40 & 10$^{2.4}$-10$^3$ & 20 & 10$^{18.2}$-10$^{18.5}$ & 10$^{7.5}$\\
& high &   80 - 200 & 10$^{3.8}$-10$^{4.1}$ & 20 &  10$^{18.2}$-10$^{18.3}$ & 10$^{7.1}$-10$^{7.2}$\\
NGC\,253%
\tablenotemark{b}
\tablenotetext{b}{\citet{2014A&A...564A.126R}; $\theta \sim$ 32''}
& low & 60 & 10$^{3.5}$ & -- & 10$^{17}$  ($N_{CO}$/$\Delta v$) & --\\
& low & 40 & 10$^{4.5}$ & -- & 10$^{17}$  ($N_{CO}$/$\Delta v$) & --\\
& high & 110 & 10$^{5.5}$  & -- & 10$^{17}$  ($N_{CO}$/$\Delta v$) & 10$^{7.5}$ (total)\\
\hline
M\,83%
\tablenotemark{c}
\tablenotetext{c}{\citet{2006AA...460..467B}} & single & 40 & 10$^{5.8}$ & -- &  10$^{18.7}$ & --\\
M\,83%
\tablenotemark{d}
\tablenotetext{d}{\citet{2001AA...371..433I}; $\theta \sim$ 20''} & low   & 30 - 150 & 10$^{2.7}$-10$^{3.5}$ & -- & 10$^{17}$-10$^{17.5}$ ($N_{CO}$/$\Delta v$) & --\\
& high &   60 - 100 & 10$^{3.5}$-10$^5$ & -- & 10$^{15.8}$-10$^{17}$ ($N_{CO}$/$\Delta v$) & 10$^{7.5}$ (total)\\
\hline
M\,82%
\tablenotemark{e}
\tablenotetext{e}{\citet{2010A&A...518L..37P}; $\theta \sim$ 46''} & high & 545 & 10$^{3.7}$ & 35 & 10$^{19}$ & 10$^{7.1}$\\
M\,82%
\tablenotemark{f}
\tablenotetext{f}{\citet{2003ApJ...587..171W}; $\theta \sim$ 40'' } & low  & 14 & 10$^{3.3}$-10$^{3.8}$ & -- &10$^{17.8}$-10$^{18.0}$ & --\\
& high & 170 & 10$^{2.8}$-10$^{3.0}$ & -- &10$^{19.5}$ & 10$^{8.2}$ (total)\\
M\,82%
\tablenotemark{g}
\tablenotetext{g}{\citet{2012ApJ...753...70K}; $\theta \sim$ 43'' } & low & 63 & 10$^{3.4}$ & -- &  10$^{18.6}$ & 10$^{7.3}$\\
& high &  447 & 10$^{4.1}$ & -- & 10$^{18.0}$ & 10$^{6.1}$\\
\hline
NGC\,4945%
\tablenotemark{h}
\tablenotetext{h}{\citet{2001A&A...367..457C}} & single & 100 & 10$^{3.5}$ & -- & 10$^{18.8}$ & --\\
NGC\,4945%
\tablenotemark{i}
\tablenotetext{i}{\citet{2008AA...479...75H}} &  (degenerate & 20 & 10$^{4.5}$ & -- & 10$^{17.9}$ & 10$^{9.1}$\\
 & solutions) & 100 & 10$^3$ & -- & 10$^{17.8}$ & $10^{9}$\\
\hline
Circinus%
\tablenotemark{h}
 & single  & 50 - 80 &10$^{3.3}$ & -- & 10$^{18.3}$ & --\\
Circinus%
\tablenotemark{i}
  & (degenerate & 20 & 10$^4$ & -- & 10$^{17.5}$ & 10$^{8.8}$\\
& solutions) & 100 & 10$^3$ & -- & 10$^{17.7}$ & 10$^{8.9}$\\
\hline
NGC\,1068%
\tablenotemark{j}
\tablenotetext{j}{\citet{2012ApJ...755...57H}; $\theta \sim$ 10''} & mid-J & 169 & 10$^{5.6}$ & 148 & -- & 10$^{6.7}$\\
& high &  571 & 10$^{6.4}$ & 269 & -- & 10$^{5.6}$\\
\hline
IC\,694%
\tablenotemark{k}
\tablenotetext{k}{\citet{2012ApJ...753...46S}} & single & 10 - 500 & $>$ 10$^{2.5}$ & -- &  10$^{18}$-10$^{18.9}$ & 10$^{8.8}$\\
\hline
NGC\,3690%
\tablenotemark{k} & single  & 10 - 1000 & $>$ 10$^{2.5}$ & -- &  10$^{18}$-10$^{18.8}$ & 10$^{8.5}$\\
\hline
Arp\,220%
\tablenotemark{l}
\tablenotetext{l}{\citet{2011ApJ...743...94R}} & low  & 50 & 10$^{2.8}$ & 1.4 &  10$^{20.3}$ & 10$^{9.7}$\\
& high & 1343 & 10$^{3.2}$ & 20 &  10$^{19.4}$ & 10$^{8.7}$\\
\hline
Mrk\,231%
\tablenotemark{m}
\tablenotetext{m}{\citet{2007ApJ...668..815P}} & low & 55 - 95 & 10$^{3}$ & -- & -- & --\\
& mid-J & 40 - 70 & 10$^{4}$ - 10$^{4.5}$ & -- & -- & 10$^{10.2}$ - 10$^{10.6}$\\
\hline
\vspace{1mm}
\end{tabular}
\end{minipage}
\label{table:prev_LVG}
References from which LVG results were extracted and the beam size for mass estimates when provided:
\end{table}

\FloatBarrier
\begin{table*}[h!]
\footnotesize
\caption{LVG Model Results: CO-to-H$_2$ conversion factor, $\alpha_{CO}$\tablenotemark{a}
\tablenotetext{a}{ $\alpha_{CO}$ is given in units of [M$_{\odot}$/(K kms$^{-1}$pc$^2$)]}}
\vspace{.5cm}
\begin{tabular}{ cccc }
& $\alpha_{CO,single-comp}$ & $\alpha_{CO,two-comp}$ & $f$\tablenotemark{b}\tablenotetext{b}{$f_{warm}$ = M$_{H_2,warm}$/M$_{H_2,total}$}$_{warm}$ [\%]\\
 \hline
 \hline
  NGC\,253 & 0.57 & 2.96 & 2.2\\
  M\,83 & 0.65 & -- & --\\
  M\,82 & 0.53 & 0.25 & 49.6\\
  NGC\,4945 & 2.43 & 4.46 & 1.7\\
  Circinus & 1.68 & 13.8 & 0.5\\
  NGC1068 & 0.67 & --& --\\
  NGC\,4418 & 0.93 & 11.8 & 85.8\\
  IC\,694 & 0.37 & 0.79 & 4.1\\
  NGC\,3690 & 0.26 & 0.87 & 18.8\\
  Arp\,220 & 0.57 & 1.52 & 2.0\\
  NGC\,6240 & 0.52 & 4.57 & 57.7\\
  Mrk\,231 & 0.64 & 0.39 & 20.5\\

 \end{tabular}
 \label{table:alpha}
\end{table*}


\end{document}